\documentclass[%
unsortedaddress,
frontmatterverbose, 
preprintnumbers,
superscriptaddress,
aip,
10pt,
showpacs,
amsmath,
amssymb,
preprint,
floatfix,
nofootinbib,
longbibliography,
onecolumn
]{revtex4-2}
\usepackage{graphicx}% Include figure files
\usepackage{dcolumn}% Align table columns on decimal point
\usepackage{bm}% bold math
\usepackage{nicefrac}
\usepackage{siunitx}
\usepackage{float}
\usepackage[citecolor=blue,linkcolor=blue,colorlinks=true,urlcolor=blue]{hyperref}
\usepackage[utf8]{inputenc}
\usepackage[T1]{fontenc}
\usepackage{etoolbox}
\usepackage{amsmath}
\usepackage{epsfig}
\usepackage{color}
\usepackage[caption=false]{subfig}
\makeatletter
\def\@email#1#2{%
 \endgroup
 \patchcmd{\titleblock@produce}
  {\frontmatter@RRAPformat}
  {\frontmatter@RRAPformat{\produce@RRAP{*#1\href{mailto:#2}{#2}}}\frontmatter@RRAPformat}
  {}{}
}%
\makeatother

\renewcommand{\vec}[1]{\mbox{\boldmath$#1$}}
\newcommand{\gafd}{Geophys. Astrophys. Fluid Dyn.}
\newcommand{\jfm}{{J. Fluids Mech.}}

\newcommand{\njp}{{New J. Phys.}}
\newcommand{\grl}{{Geophys. Res. Lett.}} 
\newcommand{\gji}{{Geophys. J. Int.}}

\begin{document}

\preprint{AIP/123-QED}

\title{
The role of magnetic boundaries in kinematic and self-consistent magnetohydrodynamic simulations of precession-driven dynamo action in a closed cylinder. 
}
% Force line breaks with \\
\author{Andr{\'e} Giesecke}
\affiliation{%
Institute of Fluid Dynamics, Helmholtz-Zentrum Dresden-Rossendorf, Bautzner Landstrasse 400, D-01328 Dresden, Germany%\\This line break forced% with \\
}%
\email{a.giesecke@hzdr.de}
\author{Mike Wilbert}%
%\email{mike.wilbert@rub.de}
\affiliation{ 
Institut f{\"u}r theoretische Physik I, Ruhr Universit{\"a}t Bochum, D-44780 Bochum, Germany%\\This line break forced with \textbackslash\textbackslash
}%
\author{J{\'a}n {\v{S}}imkanin}
\affiliation{%
Institute of Fluid Dynamics, Helmholtz-Zentrum Dresden-Rossendorf, Bautzner Landstrasse 400, D-01328 Dresden, Germany%\\This line break forced% with \\
}%
\affiliation{Institute of Geophysics, Czech Academy of Sciences, Bo{\v{c}}n{\'i} II/1401, 141 00 Prague 4 -- Spo{\v{r}}ilov, Czech Republic}
%\email{jano@ig.cas.cz}
% \altaffiliation[Also at ]{Institute of Fluid Dynamics, Helmholtz-Zentrum Dresden-Rossendorf, Bautzner Landstrasse 400, D-01328 Dresden, Germany.}
\author{Rainer Grauer}%
%\email{grauer@rub.de}
\affiliation{ 
Institut f{\"u}r theoretische Physik I, Ruhr Universit{\"a}t Bochum, D-44780 Bochum, Germany%\\This line break forced with \textbackslash\textbackslash
}%
\author{Frank Stefani}
%\email{f.stefani@hzdr.de}
 \affiliation{%
Institute of Fluid Dynamics, Helmholtz-Zentrum Dresden-Rossendorf, Bautzner Landstrasse 400, D-01328 Dresden, Germany%\\This line break forced% with \\
}%

\date{\today}% It is always \today, today,

\begin{abstract}

We numerically examine dynamo action generated by a flow of an
electrically conducting fluid in a precessing cylindrical cavity. We
compare a simplified kinematic approach based on the solution of the
magnetic induction equation with a prescribed velocity field with the
results from a self-consistent three-dimensional simulation of the
complete set of magnetohydrodynamic equations. 

In all cases, we observe a minimum for the onset of dynamo action in a
transitional regime, within which the hydrodynamic flow undergoes a
change from a large-scale to a more small-scale, turbulent
behaviour. However, significant differences in the absolute values for
the critical magnetic Reynolds number occur depending on the physical
properties of the external layers surrounding the flow active
domain. The strong influence of the electromagnetic properties of
outer layers with the large variation of the critical magnetic
Reynolds number can be related to the existence of two different
branches with dynamo action.  

In contrast to the kinematic models, the nonlinear MHD simulations
reveal a small scale dynamo solution  with the magnetic energy
remaining significantly smaller than the kinetic energy of the
flow. In irregular intervals, we observe dynamo bursts with a local
concentration of the magnetic field, resulting in a global increase of
the magnetic energy by a factor of 3 to 5. However, diffusion of the
local patches caused by strong local shear is too rapid, causing these
features to exist for only a short period so that their dynamical
impact on the dynamo remains small.  
As the magnetic field is small-scale and weak, the nonlinear feedback
on the flow through the Lorentz force remains small and arises
essentially in terms of a slight damping of the fast timescales,
whereas there is no noticeable change in flow amplitude compared to
the hydrodynamic case. 

A connection with the kinematic models can be derived by looking at
the time-averaged field of the MHD dynamo solution. This is comparable
to the eigenmode of the inefficient branch of the kinematic models,
which explains their large critical magnetic Reynolds number. 

\end{abstract}

\maketitle

%%%%%%%%%%%%%%%%%%%%%%%%%%%%%%%%%%%%%%%%%%%%%%%%%%%%%%%%%%%%%%%%%%%%%%%%%%%%

\section{Introduction}

The experimental investigation of magnetic field generation through a
flow of electrically conductive fluid is of great interest, as such a
magnetohydrodynamic dynamo process is crucial for many astrophysical
bodies and allows conclusions about their evolution and internal
structure. Following the successful first-generation dynamo
experiments in Karlsruhe \citep{stieglitz2001} and Riga
\citep{gailitis2000}, which confirmed the fundamental principle,
subsequent dynamo experiments in Cadarache (\citet{monchaux2007})
revealed a wide variety of different dynamic behaviours. Currently, a
new experiment is being set up at HZDR (\citet{stefani2015}). 
In the DRESDYN experiment, a flow of liquid sodium in a precessing cylinder is intended to generate a magnetic field. 
Precession is a repeatedly proposed driving mechanism for natural dynamos, whether for the Moon \citep{cebron2019} or for the Earth \citep{malkus1968}. 
Regarding the Lunar dynamo, which around 4 billion years ago generated
a magnetic field of comparable strength to that of today's Earth,
precession is indeed the most promising candidate to explain the
observations including the disappearance of the dynamo about 3 billion
years ago (see e.g. \citet{stys2020}). 
One of the main reasons to consider precession driven dynamo action
for the geodynamo are paleomagnetic records, which suggest that the
geodynamo was active even before the formation of the Earth's solid
inner core. This, in turn, implies that thermochemical convection,
driven by the cooling and crystallization of the core, was not always
the primary mechanism and before the onset of thermochemical
convection, alternative mechanisms must have driven the early
geodynamo. Support for this hypothesis comes from recent experiments
and numerical computations that yield rather large values for the heat
conductivity under conditions typical for the Earth's liquid core,
which imposes strict energetic constraints for the convective state,
particularly for the early geodynamo before the formation of the solid
inner Earth's core  
(see e.g. \citet{olson2013,landeau2022}).
While it is relatively clear that the laminar flow directly driven by
precession is energetically insignificant (\citet{rochester1975}), the
original idea of Malkus was based on a turbulent base state\cite{malkus1968}, which was later underpinned by estimates of an
upper limit to (viscous) dissipation by \citet{kerswell1996}. 
Indeed, precession-driven fluid flows exhibit several instabilities,
allowing both forward and backward cascades of energy transfer
(\citet{pizzi2021b}) that end up in a flow state with small-scale
turbulence superimposed by large-scale flow contributions essentially
in terms of a circulation opposite to the original rotation of the
container (\citet{wilbert2022, giesecke2024}). 
This behaviour has also been observed in the laboratory\cite{kobine1996} and so far various experiments have shown that precession is capable
of generating strong flows, principally allowing for the large flow
amplitudes required for the occurrence of the dynamo
effect\cite{manasseh1992,manasseh1994,manasseh1996,noir2001a,noir2003,mouhali2012,lin2014,lebars2022,burmann2024}. 
This contrasts with potentially more straightforward mechanisms, such as
convection, which at typical laboratory scales may not achieve
sufficiently vigorous flows due to lower driving efficiency
\cite{leorat2006,christensen2006,king2010}. 
Precession-driven dynamos also have been the subject of numerous
numerical studies, and there are now successful dynamo simulations in
all conceivable geometries, including spheres \citep{tilgner2005},
spheroids \citep{wu2009}, or cubes \citep{goepfert2016}.  
\citet{nore2011} and \citet{cappanera2016} investigated dynamo
solutions in a cylindrical geometry comparable to the DRESDYN
experiment, but with small Reynolds numbers (${\rm{Re}} \leq 4000$)
and a rather large precession ratio of ${\rm{Po}}=0.15$. The induced
magnetic fields showed a complex structure, which correspond
most closely to a quadrupolar geometry. 
Preliminary investigations with particular regard to the DRESDYN
dynamo experiment utilized a kinematic model based on the geometry of
the planned experiment and the large-scale flow from hydrodynamic
simulations (\citet{giesecke2018,giesecke2019}). These models showed
that the dynamo effect works best in the narrow transition region that
separates the subcritical from the supercritical state
(\citet{pizzi2022a,kumar2023}). 
In the present study, we deepen the investigation of the dynamo effect
in the transitional regime, investigate the influence of outer
non-fluid layers, and compare kinematic models with a self-consistent
magnetohydrodynamic (MHD) model to study the influence of the
nonlinear terms and the Lorentz force feedback on the flow.

The present study is organized as follows. In
section~\ref{sec02::kinematic}, we introduce the equations
that describe an electrically conducting fluid flow in a precessing
cylinder and the related magnetic induction. We further summarize the
essential response of the fluid to the forcing imposed by the
precession of the container and present the evolution when the forcing
is increased. The subsequent section~\ref{sec::kinematic_models} is
dedicated to the results obtained from kinematic dynamo models, where
only the induction equation is solved numerically assuming a
prescribed, time-independent flow. In this section, we detail the
impact of the electrical conductivity of the container. 
Section \ref{sec::mhd} focuses on the corresponding results obtained
from the full set of magnetohydrodynamic (MHD) equations and reveals
differences and similarities with the kinematic dynamo. 
We end our study with the conclusions in
section~\ref{sec::05_conclusions}, where we try to adopt our results
for the forthcoming dynamo experiment at HZDR. 

%%%%%%%%%%%%%%%%%%%%%%%%%%%%%%%%%%%%%%%%%%%%%%%%%%%%%%%%%%%%%%%%%%%%%%%%%%%%%%

\section{Setup of the kinematic dynamo models{\label{sec02::kinematic}}}

\subsection{Equations and numerical methods\label{sec::02a_equation_and_setup}}

Initially, we focus on the kinematic problem, where only the magnetic
induction equation is solved, and a time-independent velocity is
prescribed, which is supposed to represent the mean (time-averaged)
velocity in the experiment.
Here we compare different configurations without and with outer layers
of a conductive material surrounding the actual medium on which the
force is exerted by the precession. 
The main finding from these models is a value for the critical
magnetic Reynolds number, which is related to the minimum amplitude of
the velocity field required for a dynamo to occur. 
The kinematic approach does neither consider the back-reaction of the
magnetic field on the fluid flow via the Lorentz force nor the impact
of small-scale fluctuations that could obstruct (turbulent diffusion) or
support ($\alpha$-effect) the dynamo process. 

\begin{figure}[h!]
\subfloat[][]{\includegraphics[width=0.54\textwidth]{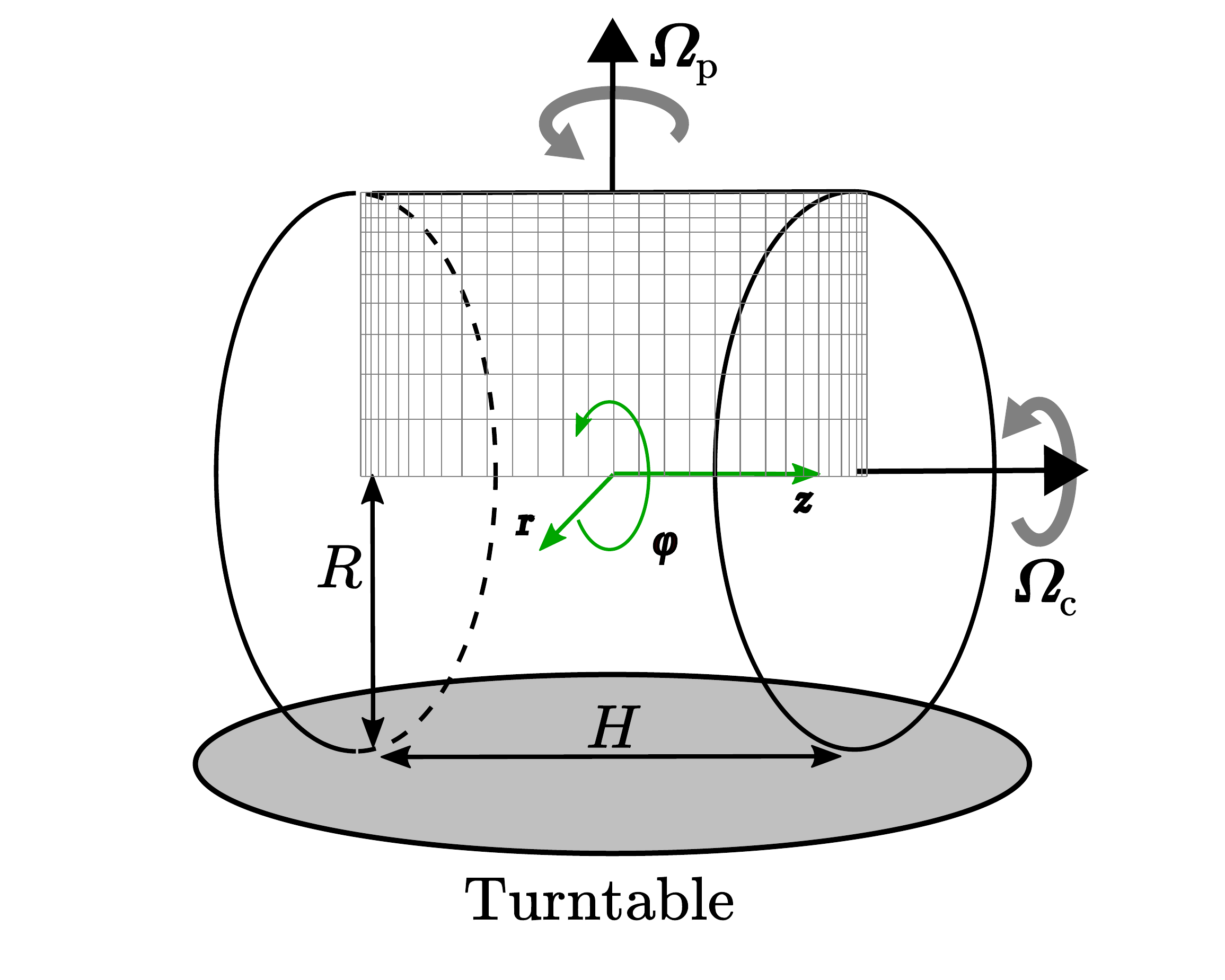}\label{fig::1a}}
%\subfloat[][]{\includegraphics[width=0.45\textwidth]{giesecke_fig_01b}\label{fig::1b}}
%\subfloat[][]{\includegraphics[width=0.45\textwidth]{poloidal_axisym_flow_with_outerlayer_thinn2}\label{fig::1b}}
\subfloat[][]{\includegraphics[width=0.45\textwidth]{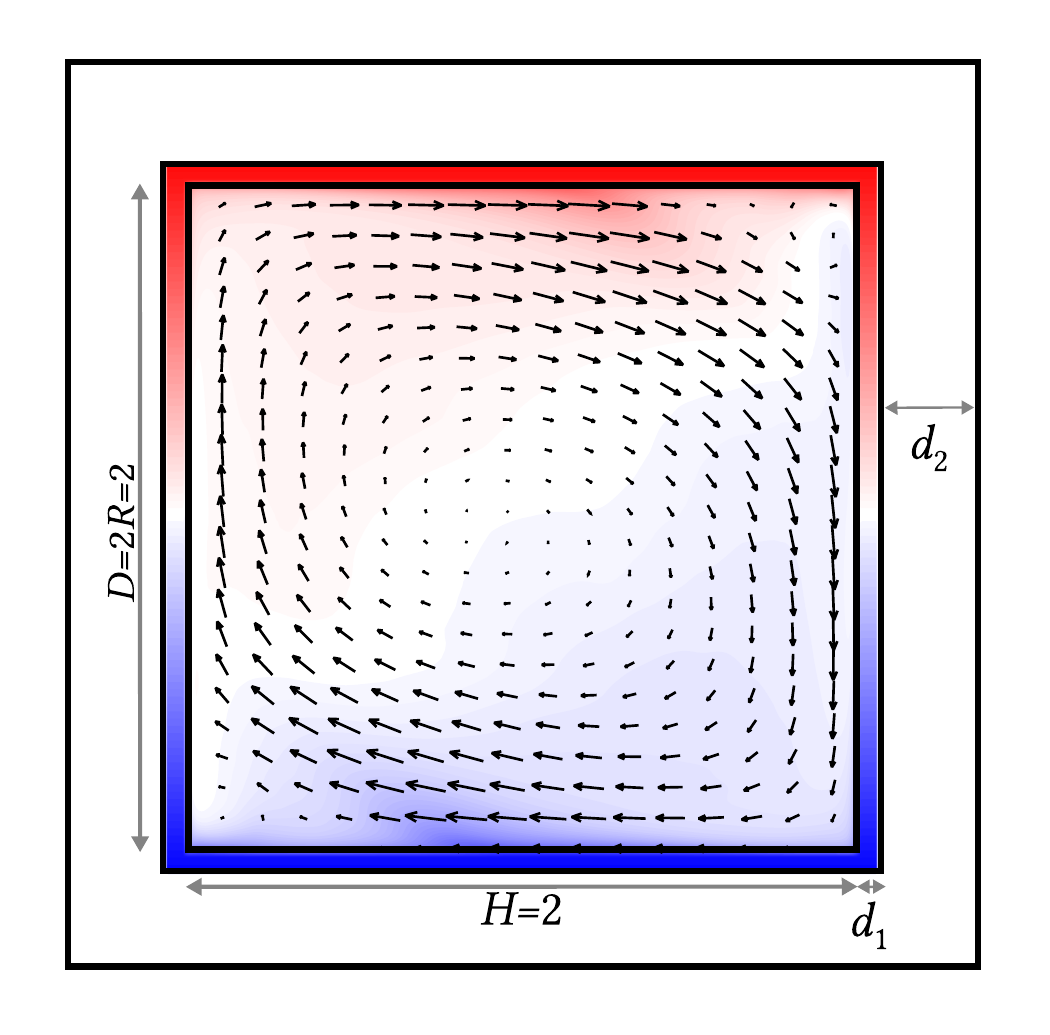}\label{fig::1b}}
\caption{
(a) Sketch of the setup labeling rotation and precession axis. The
  mesh in the meriodional plane shows the spectral elements used for
  the hydrodynamic models. 
(b) Geometric setup for the various kinematic models. The inner
  cylinder with height $H=2$ and diameter $D=2R=2$ frames the flow
  active volume, the intermediate cylinder models the container wall,
  and the very outer layer represents an outer volume with low
  electrical conductivity that resembles the laboratory exterior. The
  colors and the arrows denote a paradigmatic velocity field as
  applied in the kinematic models.\label{fig::sketch}} 
\end{figure}

We calculate the temporal evolution of the magnetic flux density
$\vec{B}$ (in the following called 'magnetic field') by numerically
solving the magnetic induction equation, which reads 
\begin{equation}
    \frac{\partial\vec{B}}{\partial t} = \nabla\times\left(\vec{u}\times\vec{B}-\frac{1}{\rm{Rm}}\nabla\times\vec{B}\right).\label{eq::induction}
\end{equation}
In Eq.~(\ref{eq::induction}) $\vec{u}$ represents the (incompressible)
velocity field and 
${\rm{Rm}}=\varOmega_{\rm{c}}R^2/\eta$ is the magnetic Reynolds number
defined with the angular velocity $\varOmega_{\rm{c}}$ of the
cylinder, the radius $R$ and the magnetic diffusivity $\eta$ 
(see also Figure~\ref{fig::1a}).
Assuming a prescribed time-independent flow field,
Eq.~(\ref{eq::induction}) specifies a linear problem and the solution
follows $\vec{B}(\vec{r},t)\propto e^{\kappa t}$ with the complex
eigenvalue $\kappa=\gamma+i\omega$ consisting of the growth rate (real
part $\gamma$) and the frequency (imaginary part $\omega$).  
In order to solve Eq.~(\ref{eq::induction}) numerically, we use a
finite volume approach in cylindrical coordinates. The scheme applies
the constraint transport method to guarantee divergence-free solutions
and is described in more detail in \citet{giesecke2008}. Extensive
tests of the algorithm, in particular the impact of locally varying
material properties, such as electrical conductivity and comparison
with an alternative approach based on Spectral/Finite Elements (the
SFEMaNS code\cite{guermond2009}) can be found in \citet{giesecke2010}.  

%%%%%%%%%%%%%%%%%%%%%%%%%%%%%%%%%%%%%%%%%%%%%%%%%%%%%%%%%%%%%%%%%%%%%%%%%%%%%%%%%%%%%%%%%%%%

\subsection{Boundary conditions and nested outer layers} 

We estimate the impact of the magnetic boundary conditions
employing various setups with one or two outer layers with thickness
$d_1$ and $d_2$ as sketched in Figure~\ref{fig::1b}.   
The underlying idea is that a sufficiently thick outer layer with poor
electrical conductivity (large magnetic diffusivity) reduces the
impact of the magnetic boundary conditions on the inner volume filled
with the electrically conductive fluid. 
We consider several models that differ in the presence of up to two
outer layers with thicknesses $d_1$ and $d_2$ that enclose the
internal region of the container filled with an electrically
conducting fluid (Figure~\ref{fig::1b}). 
In all cases, the inner region of the simulation domain consists of a
cylindrical container with radius $R=1$ and height $H=2$ (in
dimensionless units) which is filled with an electrically conducting
fluid with a magnetic diffusivity $\eta=1$.  
We examine three different geometric configurations. Initially we
revise the results for a setup that considers only the fluid interior
(i.e. $d_1=0, d_2=0$).  
The second case examines the impact of a container wall with different
values of the magnetic diffusivity (i.e. $d_1\ne 0, d_2=0,
\eta_{\rm{w}}=1\cdots 8$) and 
two different values for the thickness $d_1$ of the wall layer. In the
thin layer model we use $d_1=0.05$ with a resolution of $5$ grid cells
and in the thick layer model we use $d_1=0.25$ with a resolution of
$25$ grid cells, whereby the magnetic diffusivity in the wall layer
remains fixed at the maximum value of $\eta_{\rm{w}}=8$.  
As we tackle the problem in the precession frame of reference (with
the observer co-rotating with the precession while looking at the
spinning container), the wall layer follows a rotational motion with
$u_r=u_z=0$ and $u_{\varphi}=r\varOmega_{\rm{c}}$. 

Finally, in the fourth model we again apply a thin  wall layer with
$d_1=0.05$ and $\eta_{\rm{w}}=8$ surrounded by a second, outer layer
with  $d_2=0.25$ and $\eta_{\rm{o}}=8$, which is supposed to resemble a
non-conducting exterior (see \citet{raedler2002} and present discussion on
page 14). This layer does not co-rotate with the container so that the
velocity is set to zero within this layer. 
At the outer boundaries of the computational domain we apply
pseudo-vacuum conditions for the magnetic field, which read
($\vec{B}\times \vec{n})_{\rm{bc}}=0$, where $\vec{n}$ is the unit
normal vector at the boundary so that the tangential components of the
magnetic field an the outer boundary vanish.   
It is well known that pseudo-vacuum conditions in kinematic models
usually underestimate the critical magnetic Reynolds number that must
be exceeded for the dynamo to start. The reduction depends on the
geometry, i.e. essentially on the aspect ratio, and amounted, for
example, to up to 30\% in the kinematic models of the VKS dynamo
compared to realistic insulator boundary conditions
(\citet{giesecke2010}).

\subsection{Velocity field}

\begin{figure}[t!]
\includegraphics[width=0.65\textwidth]{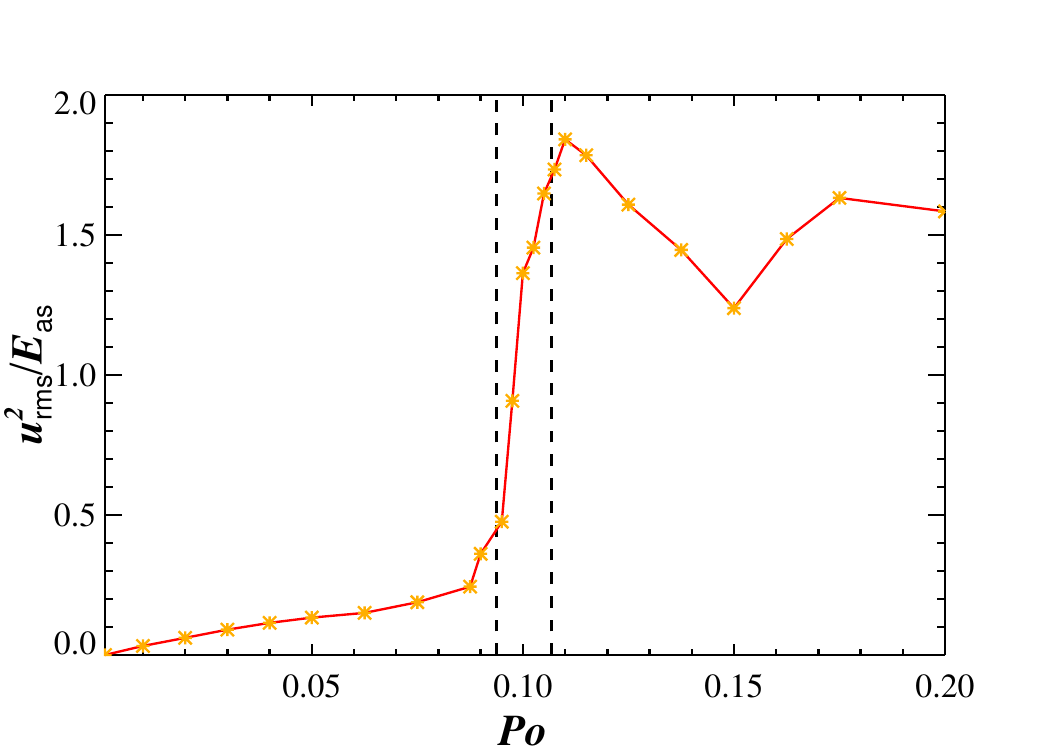}
\caption{Root mean square velocity over the kinetic energy of the
  axisymmetric geostrophic component (circulation) showing the
  transition from a rotation dominated regime to a non-rotating
  turbulence.\label{fig::transition}} 
\end{figure}

The fluid flow $\vec{u}$ is prescribed using the data taken from
hydrodynamic simulations presented in
\citet{pizzi2021a,pizzi2021b,wilbert2022} and \citet{giesecke2024}.  
In these studies, it was found that the flow in a cylinder emerging at
a large nutation angle can be represented by the time-averaged flow
with a few large-scale inertial waves that capture the major part of
the kinetic energy of the flow.  
In dependence of the precession ratio defined by
${\rm{Po}}=\varOmega_{\rm{p}}/\varOmega_{\rm{c}}$ the flow basically can be
characterized by three regimes, which we simply name  
the subcritical and the supercritical state, and the transition
section between both states. 
A compact representation of the three regimes, which illustrates how
the transition between the subcritical state and the supercritical
state takes place, is provided in Figure~\ref{fig::transition}, which
shows the ratio of the energy of the turbulent flow component in the
form of the time- and volume-averaged root-mean-square velocity  
$u_{\rm{rms}}^2=(\pi R^2H \Delta
T)^{-1}\int\left|\vec{u}(\vec{r},t)-\overline{\vec{u}(\vec{r})}\right|^2
dV dt$ to the energy of the actual axisymmetric
geostrophic component 
$E_{\rm{as}}=(\pi HR^2 \Delta
T)^{-1}\int\left|\vec{u}^{m=0}(r,z,t)\right|^2 dV dt$ as a
function of ${\rm{Po}}$. The plot illustrates the abrupt shift from
the rotation-dominated regime (below ${\rm{Po}} \approx 0.0925$) to
the turbulence-dominated regime (above ${\rm{Po}} \approx 0.1075$). 
However, although the fluctuations of the flow with respect to the
fluid rotation increase strongly with ${\rm{Po}}$, the dominant
energetic contributions of the precession-induced flow can always be
captured by a few large-scale inertial modes, since the significant
jump in Figure~\ref{fig::transition} is determined by both, the
increase in turbulence intensity, and the deceleration of the fluid
rotation in the core of the cylinder (see \citet{giesecke2024}).  

\begin{figure}[h!]
\begin{center}
  \includegraphics[width=0.99\textwidth]{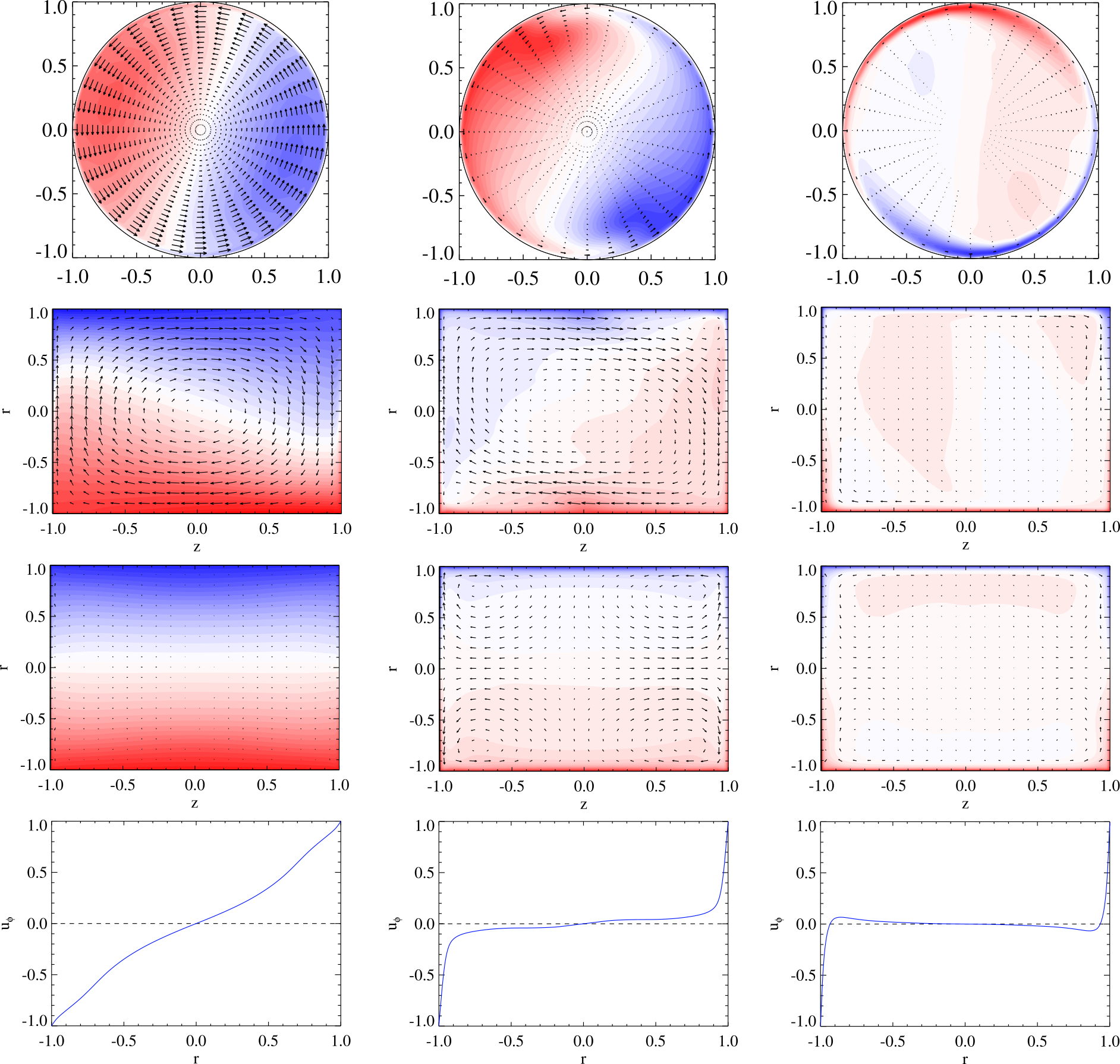}
\end{center}
\caption{Characteristic structure of the velocity field for ${\rm{Re}}=10^4$ and ${\rm{Po}}=0.03,0.1,0.2$ (from left to right). From top to bottom:
Total flow in the equatorial plane (colors: $u_z$, vectors: $u_r\vec{e}_r+u_{\varphi}\vec{e}_{\varphi}$), total flow in the meridional plane where $u_z$ is maximum (colors: $u_{\varphi}$, vectors: $u_r\vec{e}_r+u_z\vec{e}_z$), axisymmetric flow (colors: $u_{\varphi}$, vectors: $u_r\vec{e}_r+u_z\vec{e}_z$),
and the radial profile of the axisymmetric angular velocity $u_{\varphi}^{m=0}$ in the zentral plane (at $z=0$).
\label{fig::flowstructure}}
\end{figure}

The typical structure of the time-averaged flow fields is shown in
Figure~\ref{fig::flowstructure} for three precession ratios
${\rm{Po}}=0.03, 0.1, 0.2$ that are characteristic for the three flow
regimes. 
In the subcritical state (left column  in Figure~\ref{fig::flowstructure}), the flow is essentially laminar, and the geometry can be described by 
the directly driven flow (first and second row, essentially an $m=1$ mode proportional to $\cos\varphi$) and
a zonal flow maintained by the cylinder's rotation (geostrophic and
axisymmetric, third row).
Without precession, this would be a pure solid-body rotation 
$u_{\varphi}\hat{\vec{e}}_{\varphi}=f(r)\hat{\vec{e}}_{\varphi}$ with $f(r)=r\varOmega_{\rm{c}}$ but due to the influence of precession 
the radial profile $f(r)$ of the zonal flow is significantly modified
(i.e. deceleration of the fluid rotation in the bulk, see bottom row
in Figure~\ref{fig::flowstructure}).  
For larger forcing the flow transients into a supercritical region, in
which the large-scale flow in the inner region almost vanishes so that
the fluid essentially performs small-scale turbulent fluctuations and
large-scale components are only found near the walls of the container 
(see right column in Figure~\ref{fig::flowstructure}). 
These two regimes are linked by a transition zone, which is
characterized by an intermittent behavior of the large-scale flow with
elements from both regimes.  
In this transitional region, there exists an additional large-scale
contribution in the form of a double roll pattern as shown in the
third row of the central column of Figure~\ref{fig::flowstructure}. 

This double roll is rather similar to the (time-averaged) poloidal
velocity field in the VKS dynamo\footnote{In that dynamo experiment
the flow was driven by two oppositely rotating disks}, and it is well known that such a flow structure generates a
dynamo with a relatively low critical magnetic Reynolds number\cite{dudley1989}  (when also the associated azimuthal flow is taken into account).
Indeed, previous kinematic models using a time-averaged flow
    driven by precession demonstrated that dynamo action is most
effective when utilizing the flow obtained in the transitional region
between the subcritical and supercritical states
\cite{giesecke2018,giesecke2019}.  
A detailed investigation of which components (in terms of inertial
modes) are of decisive importance for the dynamo effect was carried
out in \citet{giesecke2019} and showed that 
shear adjacent to the side walls plays an important role as well as
the presence of the double-roll mode which seems to be responsible for
the considerable drop in the critical magnetic Reynolds number.

\section{Kinematic growth rates and critical magnetic Reynolds number\label{sec::kinematic_models}}

\subsection{General overview}

So far kinematic models for the DRESDYN dynamo experiment mostly
applied pseudo-vacuum conditions for the magnetic field and only few
models have been published where an outer wall with finite thickness
and with different magnetic diffusivity has been considered
\cite{giesecke2019}. These models indicate that
a container wall made of stainless steel, i.e. with a diffusivity that
is larger by one order of magnitude when compared to the liquid sodium
in the interior, yields growth rates for the magnetic energy similar
to the models with pseudo-vacuum conditions. In the following, we will
review and deepen these results. 

\subsection{No layer}

\begin{figure}[t!]
\begin{center}
\subfloat[][]    
{\begin{minipage}{0.68\textwidth}
%{\vspace*{-0.3cm}\includegraphics[width=\textwidth]{magener_vs_tim_re10000_po0p10_hires.pdf}}
%{\vspace*{-0.5cm}\includegraphics[width=\textwidth]{magener_vs_tim_re10000_po0p10_po0p125.pdf}}
{\vspace*{-0.5cm}\includegraphics[width=\textwidth]{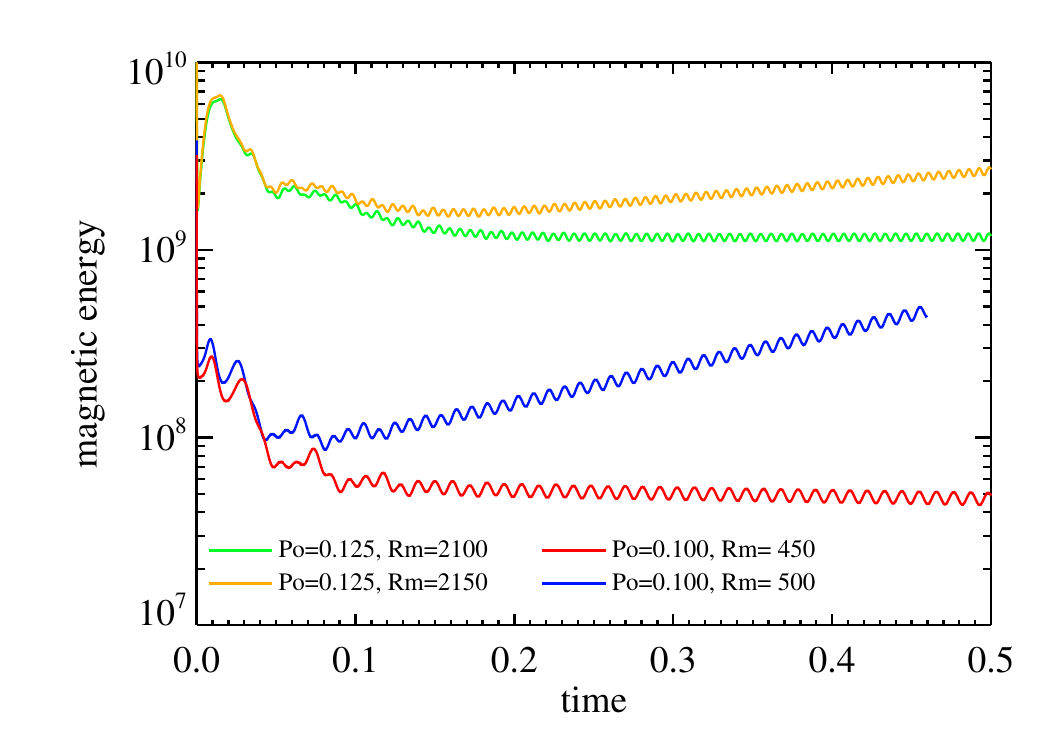}}
\end{minipage}\label{fig::magener}}
 \subfloat[][]    
%{\includegraphics[width=0.32\textwidth]{emag_kinematic_nolayer.png}\label{fig::3b}}
{\includegraphics[width=0.32\textwidth]{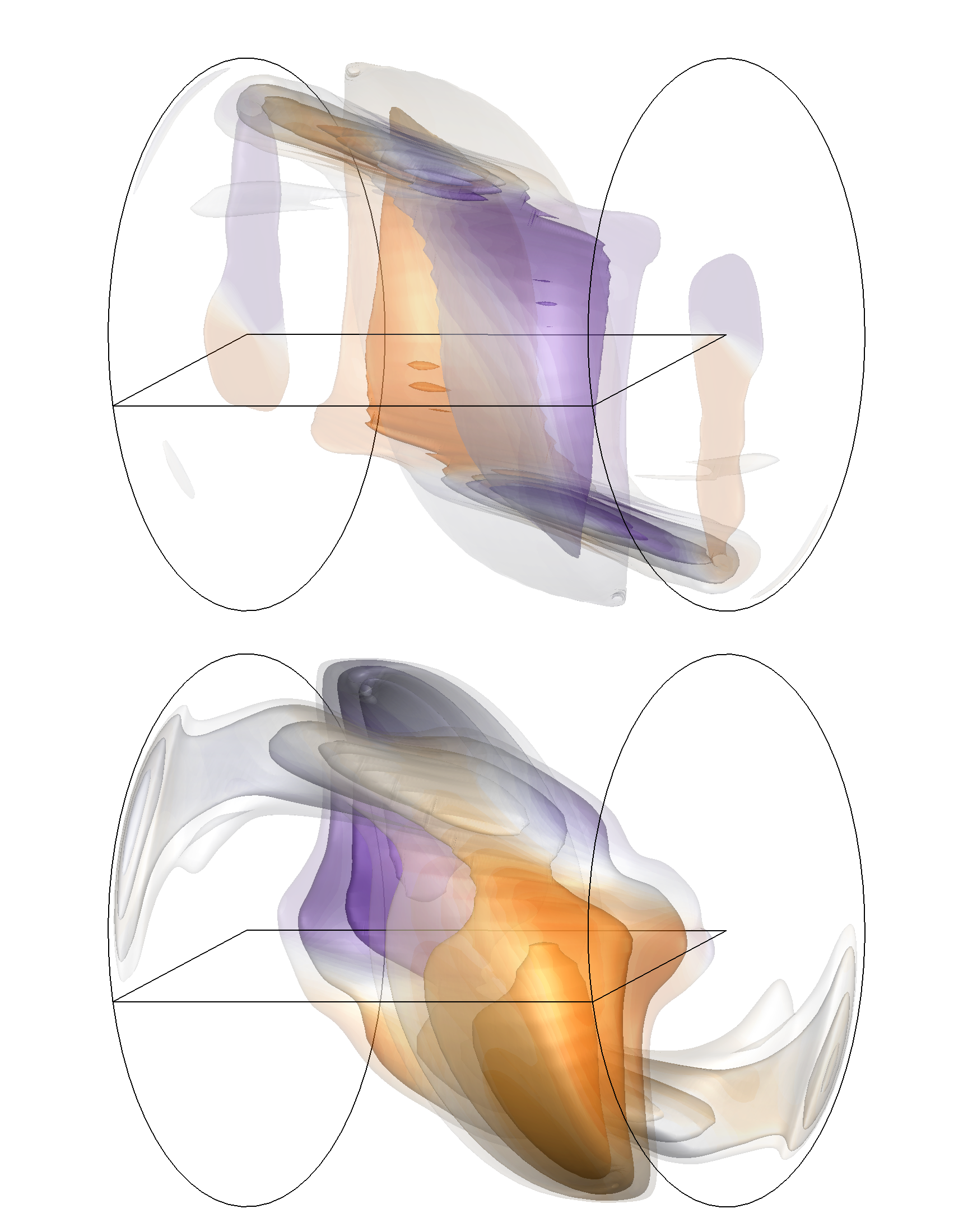}\label{fig::3b}}
\end{center}
\caption{
\label{fig::ener_nolayer}
(a) Magnetic energy versus time for two particular cases with
${\rm{Rm}}=450$ (red) and ${\rm{Rm}}=500$ (blue) obtained with the
flow field at ${\rm{Po}}=0.1$ 
and for ${\rm{Rm}}=2100$ (green) and ${\rm{Rm}}=2150$ (orange)
obtained with the flow field at ${\rm{Po}}=0.125$. 
(b) Nested iso-surfaces of the magnetic energy at $15\%, 30\%$ and
$50\%$ of the maximum value for the flow field taken at
${\rm{Po}}=0.10, {\rm{Rm}}=500$ (bottom)  
and for the flow field taken at ${\rm{Po}}=0.125$ and ${\rm{Rm}}=2100$
(top). Both plots are snapshots from an oscillatory state and the
corresponding animation movie can be found in the supplemental
material of this study\cite{supplement1}. 
}
\end{figure}
The typical behavior of the temporal evolution of the magnetic energy 
$E_{\rm{m}}=1/2\int \vec{B}^2dV$ is shown in
Figure~\ref{fig::magener}. Here we focus on two paradigmatic cases
with the flow obtained at ${\rm{Po}}=0.1$ scaled to a magnetic
Reynolds number of ${\rm{Rm}}=450$ (red curve, decaying solution) and
${\rm{Rm}}=500$ (blue curve, growing solution), which are slightly
below and above the dynamo threshold.  
A striking feature is the oscillating component of the energy, which is
superimposed on the exponentially growing part.  
For a flow field with a slightly increased ${\rm{Po}}$, the behavior
looks similar in principle, but then a higher magnetic Reynolds number
is required for the onset of a dynamo, and the frequency of the
oscillating component increases as shown by the orange and green
curves in Figure~\ref{fig::magener}, which result from the flow field
obtained at ${\rm{Po}}=0.125$. 

The distribution of the magnetic energy looks similar in both cases
(see Figure~\ref{fig::3b}). It is worth mentioning in relation to
results including outer boundary layers (see below) that a significant
part of the magnetic field energy is observed in the bulk of the
cylinder. The temporal dynamics due to the oscillating field component
are shown in an animation, which can be found in the supplementary
material to this study \cite{supplement1}. 

\subsection{Wall layer}

We continue with the setup that considers an additional wall layer with thickness $d_1=0.05$. Within this layer the magnetic diffusivity $\eta_{\rm{w}}$ is increased in order to emulate the properties of a container wall made of stainless steel. With such an outer layer, we essentially find two new features: (1) the onset for dynamo action is shifted to larger magnetic Reynolds numbers 
\begin{figure}[t!]
      \subfloat[][]{\includegraphics[width=0.68\textwidth]{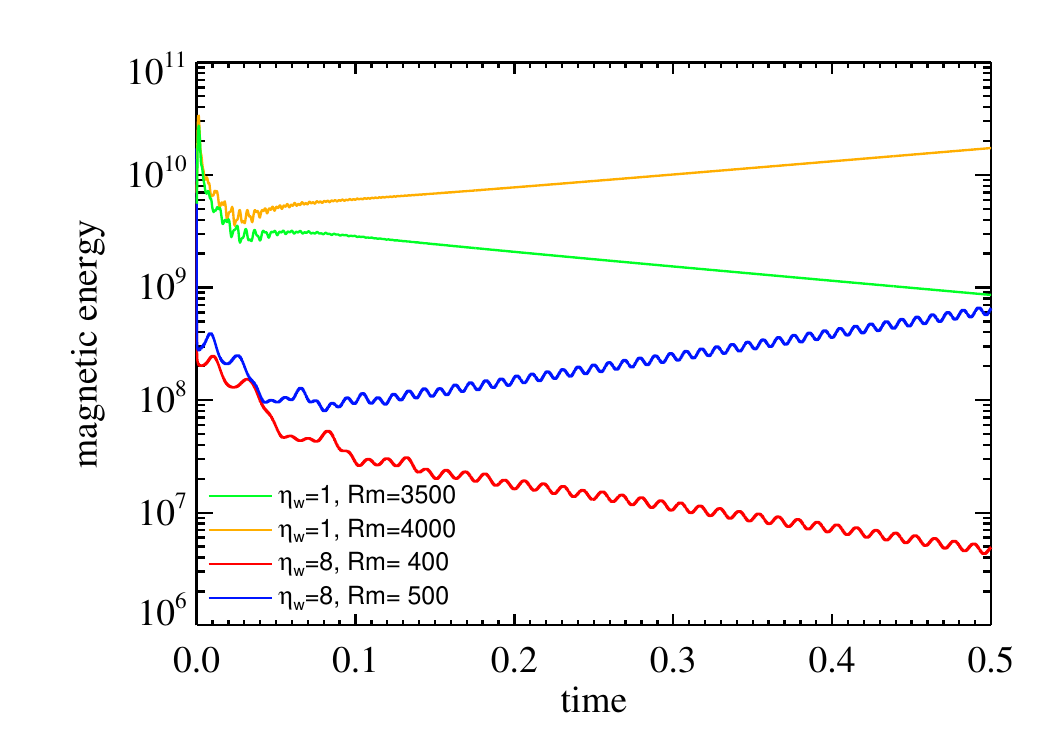}
    \label{fig::ener_vs_tim_thinlayer}}
%      \subfloat[][]{\includegraphics[width=0.48\textwidth]{giesecke_fig_04b}\label{fig::gr_thinlayer_var_eta}}
%     %\includegraphics[width=0.45\textwidth]{gr_vs_rm_var_eta_detail.pdf}
%     \\[-0.5cm]
%\subfloat[][]{
%\begin{minipage}{0.4\textwidth}
%\includegraphics[width=0.99\textwidth]{giesecke_fig_04c.png}%%\label{fig::isosurface_thinlayer_eta1p0}   
%\\[-1cm]
%\includegraphics[width=0.99\textwidth]{giesecke_fig_04d.png}%%\label{fig::isosurface_thinlayer_eta8p0}
%\end{minipage}\label{fig::thicklayer}}
\subfloat[][]{\includegraphics[width=0.32\textwidth]{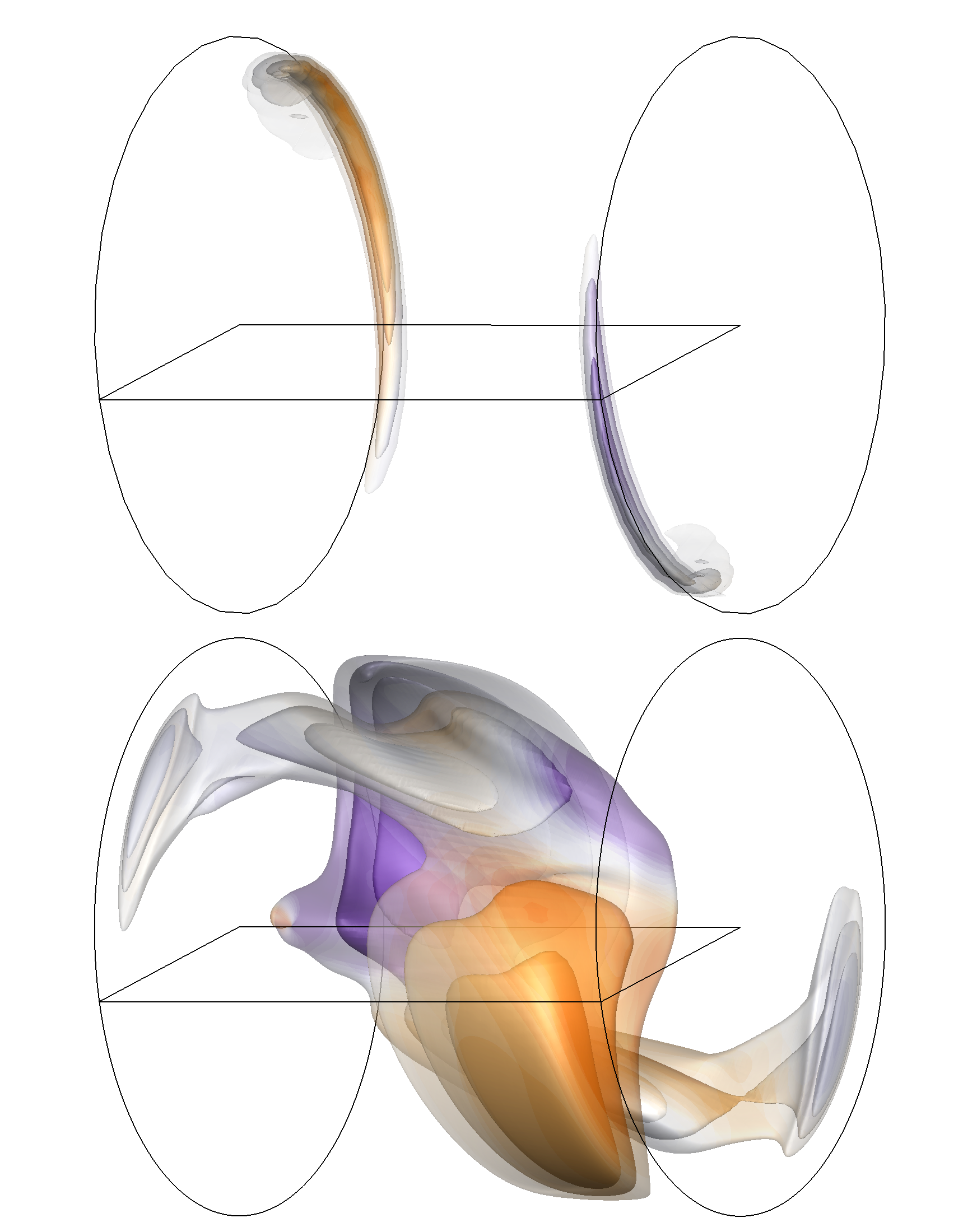}
    \label{fig::ener_with_layer}}
\caption{\label{fig::growthrates}
(a) Magnetic energy versus time for thin outer layer with
  $\eta_{\rm{w}}=1$ (same as fluid, green and orange) and
  $\eta_{\rm{w}}=8$ (red and blue). In both cases the evolution around
  the onset of dynamo action is shown for the flow field obtained at
  ${\rm{Po}}=0.1$.  
(b) Distribution of magnetic energy for a thin layer solution with
  $\eta_{\rm{w}}=1$ and ${\rm{Rm}}=4000$ (top) corresponding to the
  orange curve in plot (a) and for a thin layer with $\eta_{\rm{w}}=8$
  and ${\rm{Rm}}=500$ corresponding to the blue curve in plot (a). The
  nested isosurfaces represent the magnetic energy at $15\%, 30\%$ and
  $50\%$ of the maximum value and the colored mapping denotes the
  radial field $B_r$. Note that the bottom solution exhibits a
  oscillating behaviour similar to the solutions shown in
  Figure~\ref{fig::3b}.  
}
\end{figure}
and (2) we can distinguish two different types of dynamo action: if
the wall diffusivity is close to that of the fluid
(i.e. $1<\eta_{\rm{w}}\lesssim{2}$), a dynamo only occurs at very
large ${\rm{Rm}}$ (${\rm{Rm}}^{\rm{crit}} > 3500$). This dynamo has no
oscillatory components (see orange and green curve in
Figure~\ref{fig::ener_vs_tim_thinlayer}) and the associated field
energy is concentrated in a narrow region near the end
caps (top plot with iso-surfaces in
Figure~\ref{fig::ener_with_layer}). 
If the wall layer has an $\eta_{\rm{w}} > 2$ (see red and blue curves
in Figure~\ref{fig::ener_vs_tim_thinlayer}), we obtain an oscillating
solution that is similar to the solution in the previous case with
pseudo-vacuum boundary conditions. 
In that case the magnetic field energy is again widely distributed
throughout the entire volume with the characteristic pattern shown in
the bottom plot of Figure~\ref{fig::ener_with_layer}. 
\begin{figure}
   \includegraphics[width=0.65\textwidth]{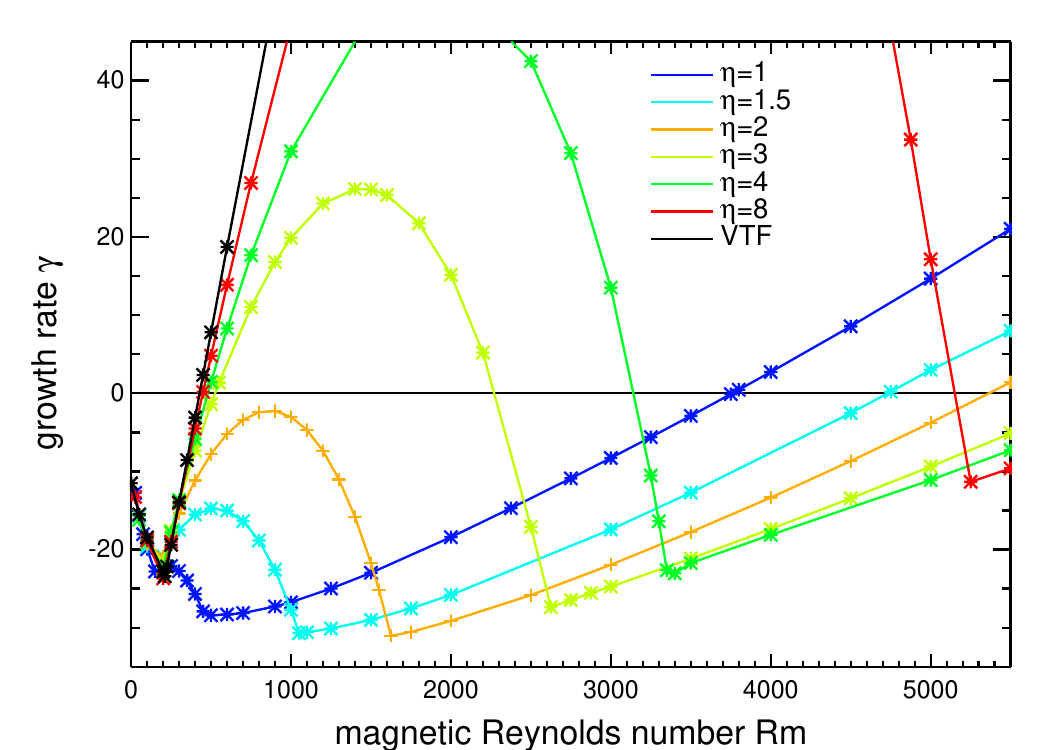}
\caption{\label{fig::gr_thinlayer_var_eta}
   Growth rates $\gamma$ of the magnetic field for the time-averaged
   flow obtained at hydrodynamic simulations at ${\rm{Re}}=10^4$ and
   ${\rm{Po}}=0.1$ with various values of the magnetic diffusivity in
   the thin container walls ($d_1=0.05$).  
   The black curve denotes $\gamma$ for setups with vanishing
   tangential fields being imposed without container walls ($d_1=0$). 
}
\end{figure}

%%%%%%%%%%%%%%%%%%%%%%%%%%%%%%%%%%%%%%%%%%%%%%%%%%%%%%%%%%%%%%%%%%%%%%%%%%%%%%%%%%%

\begin{figure}[b!]
\subfloat[][]{\includegraphics[width=0.51\textwidth]{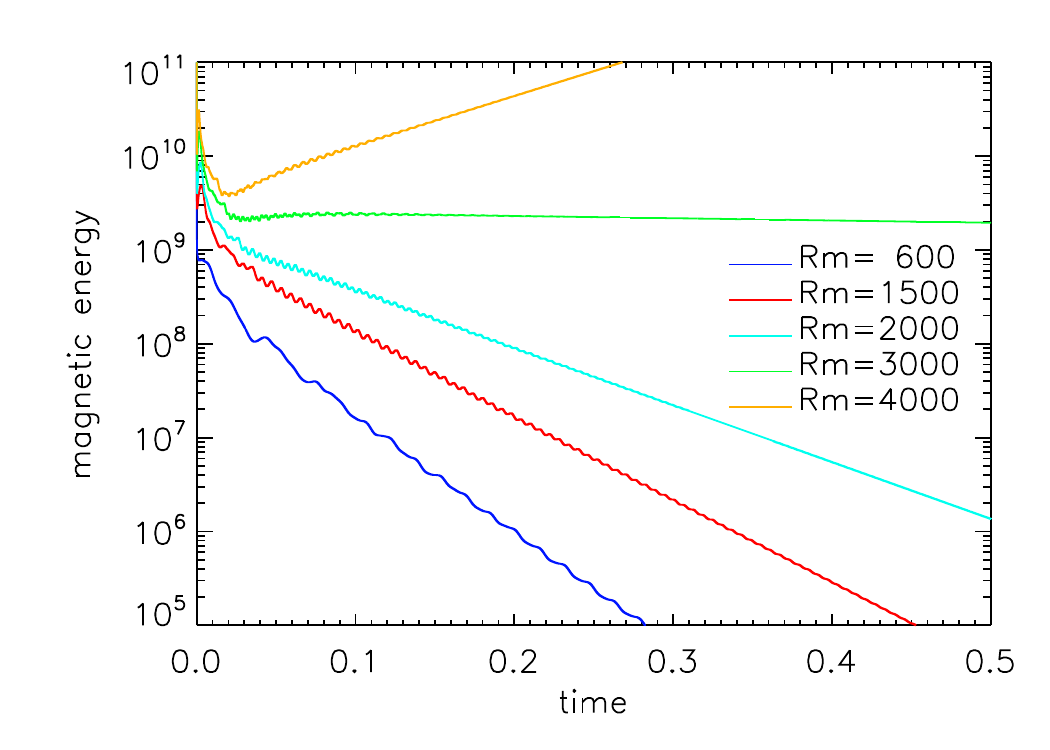}\label{fig::ener_vs_tim_thicklayer}}
%\subfloat[][]{\includegraphics[width=0.49\textwidth]{giesecke_fig_05b}\label{fig::ener_vs_tim_two_layers}}
%    \\
\subfloat[][]{\includegraphics[width=0.46\textwidth]{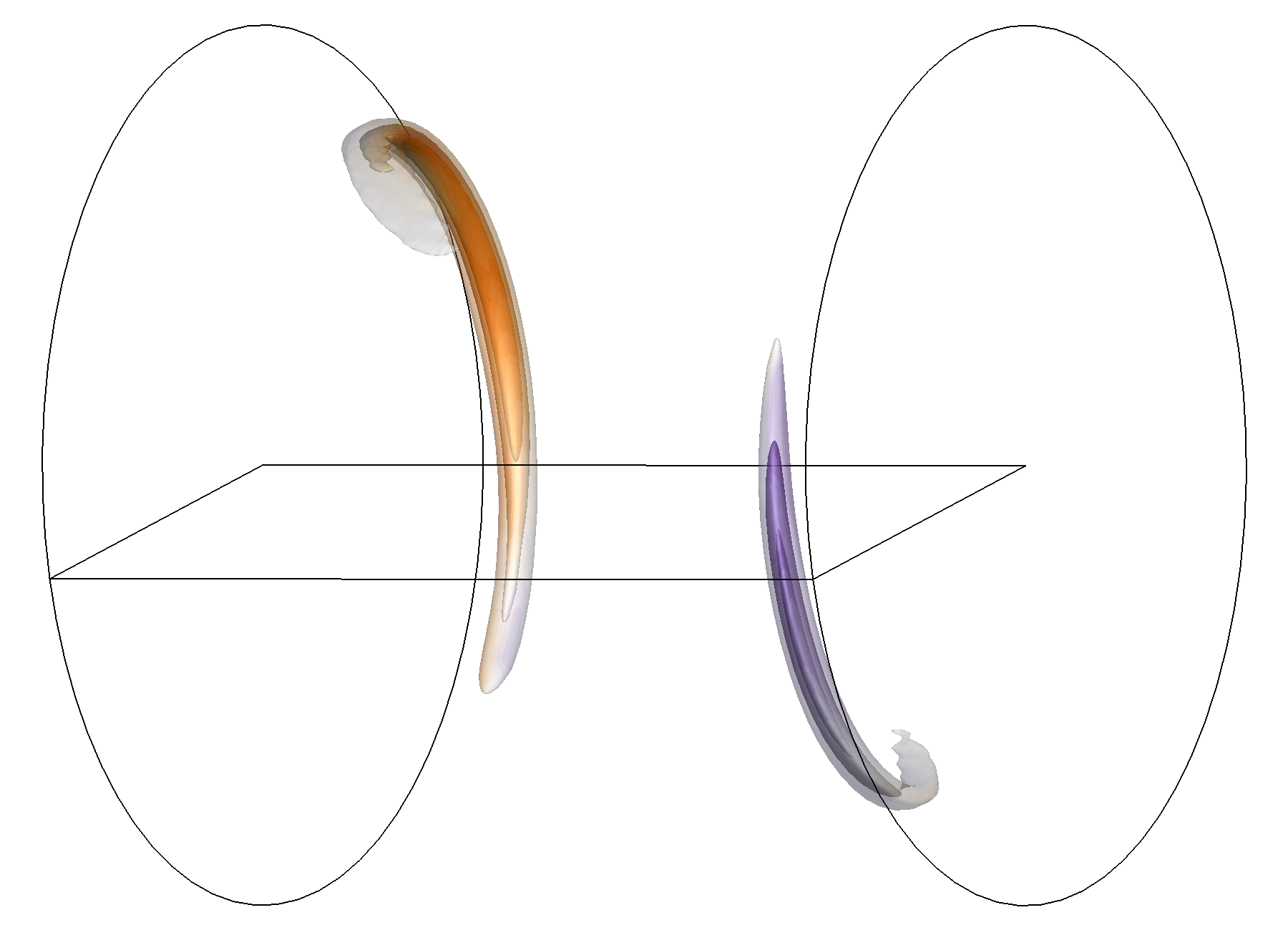}\label{fig::structure_thicklayer}}
%\subfloat[][]{\includegraphics[width=0.49\textwidth]{giesecke_fig_05d.png}\label{fig::structure_thinlayer_with_outer_region}}
\caption{\label{fig::ener_vs_time_thicklayer}
(a) Magnetic energy versus time for the setup with thick container wall ($d_1=0.25, d_2=0$). 
(b) isosurfaces for thick layer with $\eta_{\rm{w}}=8, {\rm{Rm}}=3000$ (green curve in (a)), 
}
\end{figure}
A detailed view on the impact of the diffusivity of the outer layer on
the growthrates is shown in Figure~\ref{fig::gr_thinlayer_var_eta}.  
When the magnetic diffusivity of the outer layer is increased to
emulate the effect of a stainless steel container, an eigenmode is
excited, the growth rates of which are represented by the inverted
parabola in the individual colored curves in
Figure~\ref{fig::gr_thinlayer_var_eta}. For $\eta_{\rm{w}} \gtrsim 2$
this special shape results in positive growth rates at  ${\rm{Rm}}$
far below the point where the "regular" curve becomes positive. For
further increasing diffusivities the behaviour approaches that of
pseudo-vacuum boundary conditions without outer layer (illustrated by
the black curve).  
It is therefore the eigenmode that belongs to the inverse parabolic
growth rate, which corresponds to a dynamo that occurs at relatively
low ${\rm{Rm}}$ and causes a magnetic field that permeates the entire
volume.  

%%%%%%%%%%%%%%%%%%%%%%%%%%%%%%%%%%%%%%%%%%%%%%%%%%%%%%%%%%%%%%%%%%%%%%%%%%%%%%%%%%%%%%%%%%%%%%%%

The beneficial effect of the wall layer with (sufficiently) large
$\eta_{\rm{w}}$ vanishes, when its thickness $d_1$ is increased, as
shown in Figure~\ref{fig::ener_vs_tim_thicklayer} which presents the
evolution of the magnetic energy for $d_1=0.25$ while the magnetic
diffusivity remains at $\eta_{\rm{w}}=8$. Here we end up with a
critical magnetic Reynolds number slightly above
${\rm{Rm}}^{\rm{crit}}\approx 3000$ and the solution is again
characterized by a non-oscillatory behavior with the magnetic energy
constraint to two small section close to the endcaps (see
Figure~\ref{fig::structure_thicklayer}), indicating that we are in the
low efficiency branch. 

\subsection{Wall layer plus outer layer}

\begin{figure}[h!]
\subfloat[][]{\includegraphics[width=0.51\textwidth]{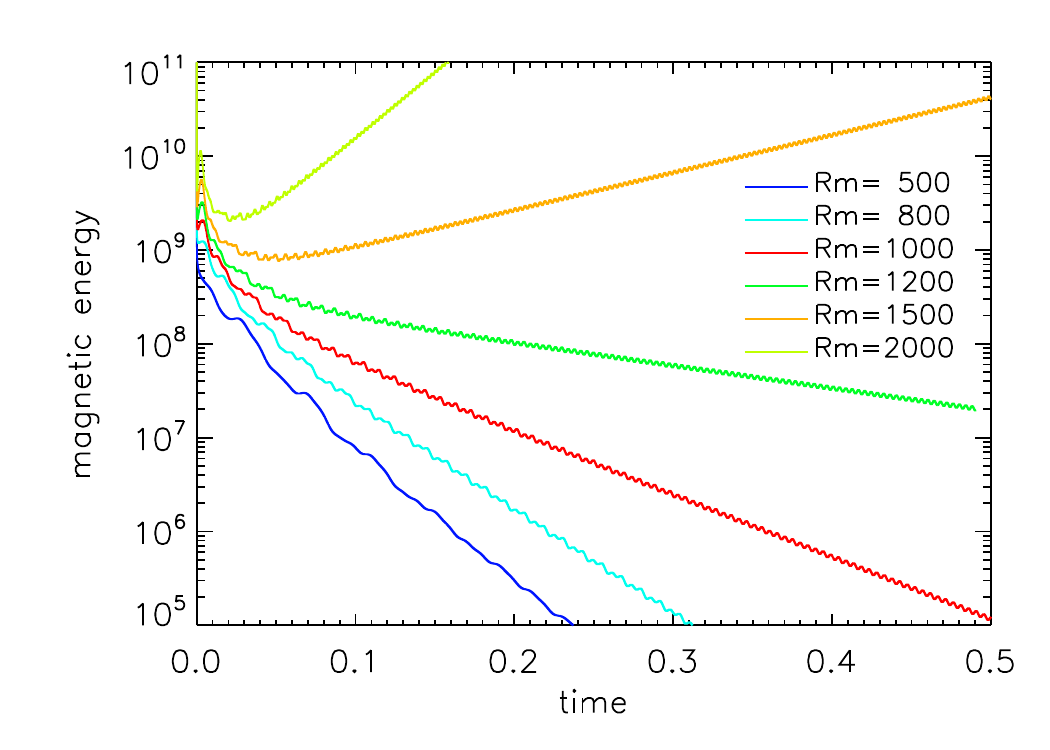}\label{fig::ener_vs_tim_two_layers}}
\subfloat[][]{\includegraphics[width=0.46\textwidth]{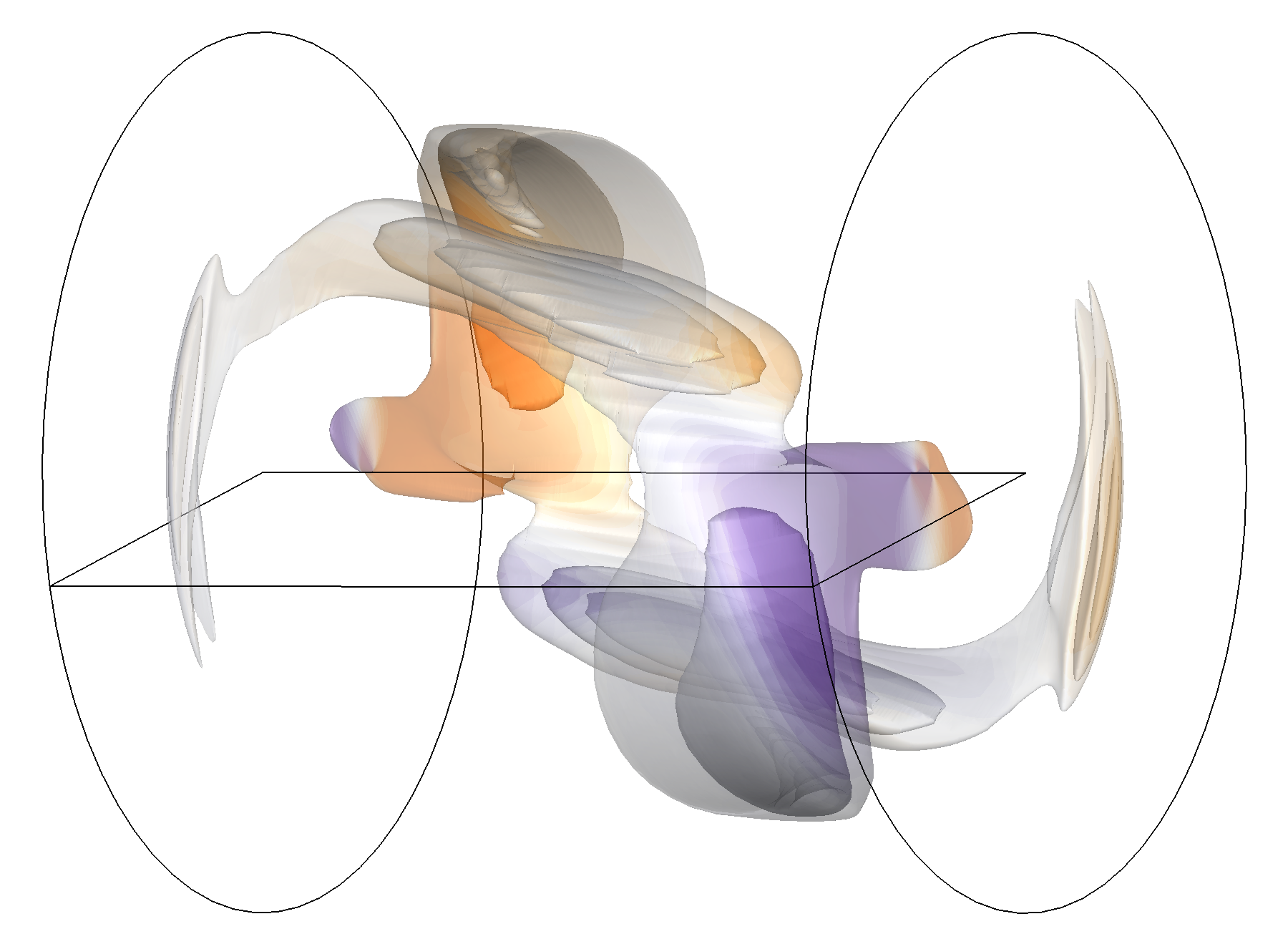}\label{fig::structure_thinlayer_with_outer_region}}
\caption{\label{fig::ener_vs_time_thicklayer2}
(a) Magnetic energy versus time for the setup with thin outer wall and
  stagnant outer layer ($d_1=0.05, d_2=0.25$) and  
(b) thin wall with $\eta_{\rm{w}}=8, {\rm{Rm}}=1500$, outer resting
  layer (orange curve in (a)).  
}
\end{figure}

The high-efficient branch can also be achieved by assuming a
configuration, where the container wall is emulated by a thin
co-rotating layer with $d_1=0.05$, while the exterior domain  
%(i.e. the {\it{laboratory}}) 
is modeled by a thick resting outer layer with $d_2=0.25$. Here both
additional layers have the same diffusivity,
i.e. $\eta_{\rm{w}}=\eta_{\rm{o}}=8$ (actually, since the thick outer
layer emulates the exterior this should be an insulator with
$\eta_{\rm{o}}\rightarrow \infty$). In this case, we again obtain an
oscillating solution that occupies the entire cylinder volume (see
Figure~\ref{fig::ener_vs_time_thicklayer2}) and has a similar
structure as in the case of the thin layer or pseudo-vacuum boundary
conditions without an outer layer shown in
Figures~\ref{fig::ener_nolayer} and \ref{fig::growthrates}. However,
the critical magnetic Reynolds number of ${\rm{Rm}}^{\rm{crit}}
\approx 1350$ is now about two and a half times as large as in the
previous cases and the amplitude of the oscillating part remains
rather small. 

\subsection{Impact of the structure of the velocity field}

\begin{figure}[b!]
%   \subfloat[][]{\includegraphics[width=0.45\textwidth]{giesecke_fig_04b}\label{fig::gr_thinlayer_var_eta}}
{\includegraphics[width=0.75\textwidth]{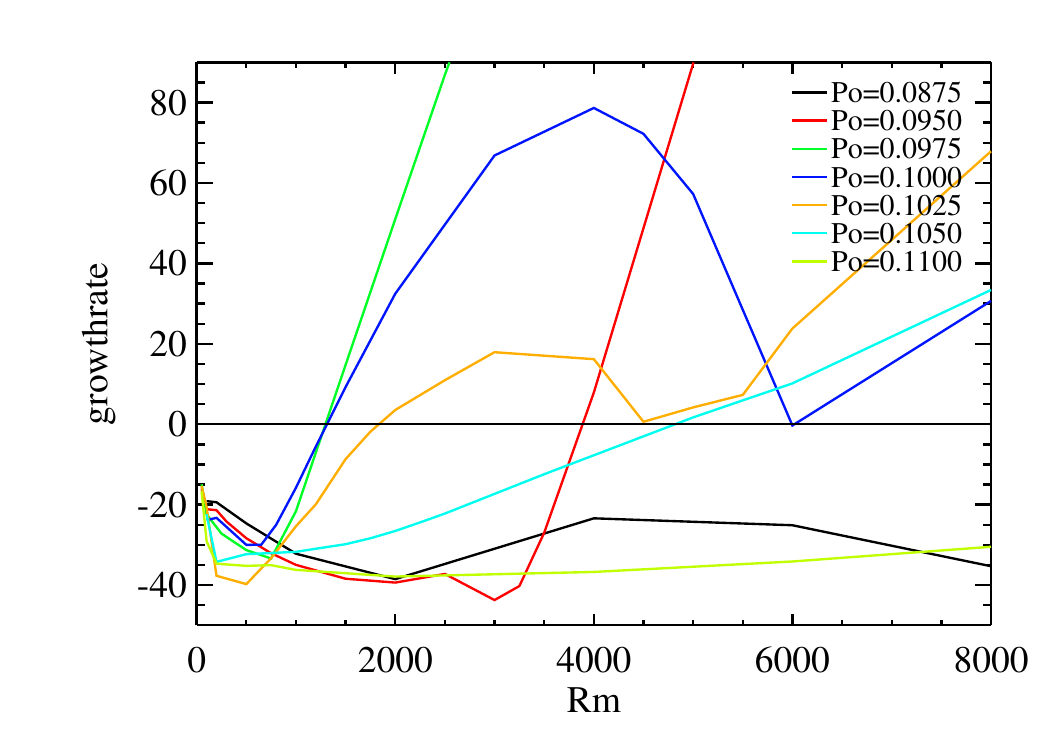}}
     %\includegraphics[width=0.45\textwidth]{gr_vs_rm_var_eta_detail.pdf}
     %\includegraphics[width=\textwidth]{giesecke_fig_03b}\label{fig::rmcrit}
%    \quad
%   \subfloat[][]{\includegraphics[width=0.52\textwidth]{rmc_vs_eta.pdf}\label{fig::rmcrit_vs_eta}}
\caption{\label{fig::fig08}
   Growth rates for the setup with wall layer and outer layer for
   various realisations of the velocity field obtained at different
   ${\rm{Po}}.$ 
%{\red{PLOT WILL BE UPDATED...Simulations are running}}
}
\end{figure}

In the following we focus on the setup with wall layer and outer layer
for which we applied various time averaged velocity fields obtained
within the interval $0.0875\leq {\rm{Po}}\leq 0.11$. 
It is known from the hydrodynamic investigations that the structure of
the velocity field changes only slightly within the transition between
subcritical and supercritical state. Nevertheless, these gradual
changes are sufficient to cause a variation in the critical magnetic
Reynolds number, especially at the borders of the transition region. 
Figure~\ref{fig::fig08} shows the growth rates for the model with wall
and outer layers but varying the flow structure by using time-averaged
flow fields obtained at different ${\rm{Po}}$.  
Similar to the results with wall layer only (Figure~\ref{fig::gr_thinlayer_var_eta}), we see a behavior for some velocity fields that can be described by an inverse parabola, at least around the range with positive growth rates. This does not apply to velocity fields with ${\rm{Po}} \leq 0.0925$ or with ${\rm{Po}} \geq 0.1075$. For these parameters, the parabolic behavior does not occur, and positive growth rates are found  - if at all - for very large ${\rm{Rm}}$. 

\begin{figure}
 \includegraphics[width=0.62\textwidth]{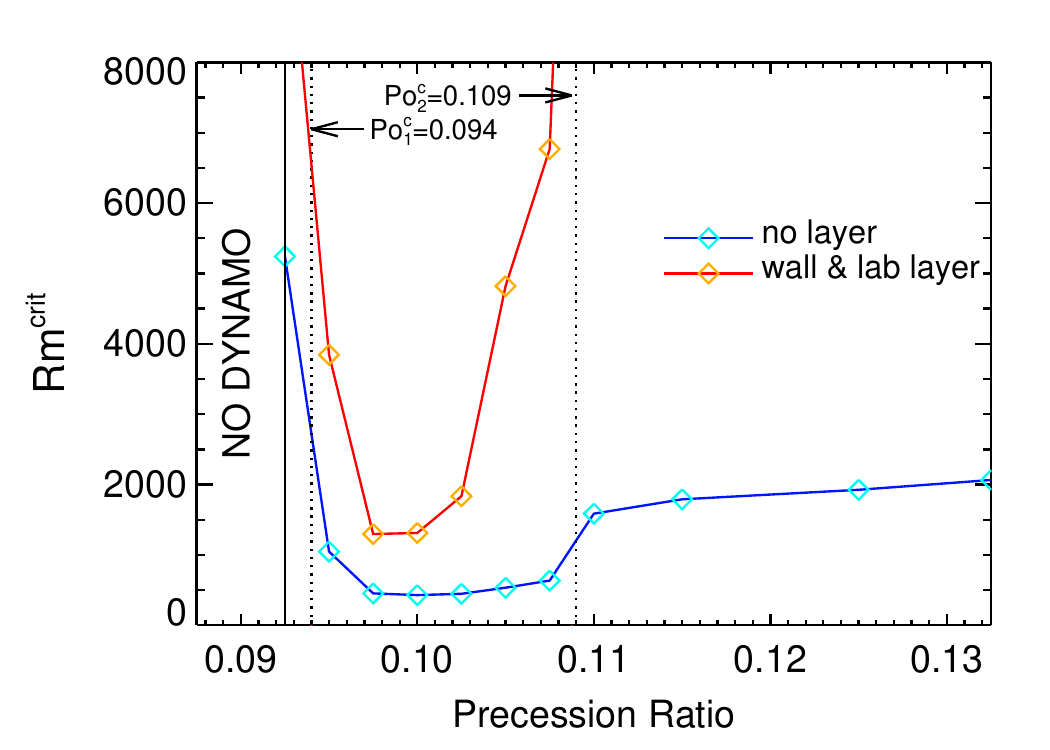}
   \caption{\label{fig::rmcrit_vs_po}
    Critical magnetic Reynolds number required for the onset of dynamo
    action using time-averaged flow fields from hydrodynamic
    simulations at ${\rm{Re}}=10^4$. The dotted vertical lines denote
    the transitional section (with a significant contribution of the
    double-roll mode). The blue curve shows the case $d_1=0$ (no
    layer) and the red curve shows the case $d_1=0.05, d_2=0.25$ (2
    outer layers).} 
\end{figure}

The intersection of the growth rates with the abscissa reveals the
critical magnetic Reynolds number required for the onset of a
dynamo. Figure~\ref{fig::rmcrit_vs_po} shows this number as a function
of ${\rm{Po}}$ for two different setups. The blue curve shows
${\rm{Rm}}^{\rm{crit}}$ for the original setup without any outer
layer.  
The most outstanding feature of the blue curve
is the reduction of the critical magnetic Reynolds number
in the range $0.095\lesssim {\rm{Po}} \lesssim 0.11$ 
with the lowest value being ${\rm{Rm}}^{\rm{crit}}\approx 450$,
obtained for the flow field from hydrodynamic simulations at
${\rm{Po}}=0.1$. Below ${\rm{Po}}=0.0925$ no more dynamos were found,
while above ${\rm{Po}}=0.11$ the critical magnetic Reynolds number is
in the range of about ${\rm{Rm}}^{\rm{crit}} \approx 2000$.  
The region $0.095\lesssim {\rm{Po}} \lesssim 0.11$ coincides exactly
with the transition region from the subcritical to the supercritical
state of the velocity field, and we therefore assume that the axially
symmetric double roll occurring exclusively in this parameter range
plays a significant role for the onset of dynamo action at
comparatively small magnetic Reynolds numbers (see
\citet{giesecke2019}).  

This behaviour changes, when considering the case with two
outer layers that is represented by the red curve in
Figure~\ref{fig::rmcrit_vs_po}. In these models the  
onset for dynamo action is shifted to larger magnetic Reynolds numbers
so that the minimum is now ${\rm{Rm}}^{\rm{c}}\approx 1300$ which is
found for the flow field obtained at ${\rm{Po}}=0.0975$. 
Furthermore, the range of flow fields with a (relatively) small
critical magnetic Reynolds number narrows and only spans the interval
$0.095 \lesssim {\rm{Po}} \lesssim 0.105$. Interestingly, although we
still use the same structure for the flow fields, we get a large
${\rm{Rm}}^{\rm{c}}$ even when the double roll mode is present with a
significant amplitude. This indicates that the impact of the double
roll mode might be less significant than expected from the cases
without outer layers but that the electromagnetic coupling of the
field induced essentially by shear close to the fluid boundary layers
with the side walls with finite electrical conductivity has a
similarly large influence. 

%%%%%%%%%%%%%%%%%%%%%%%%%%%%%%%%%%%%%%%%%%%%%%%%%%%%%%%%%%%%%%%%%%%%%%%%%%%%%%%%%%%%%%%%%%%%%%%%

\section{Full set of magnetohydrodynamic equations\label{sec::mhd}}

In the following we present results obtained from the simulation of
the induction equation coupled with the Navier-Stokes equation in a
self-consistent MHD model. 

\subsection{Code scheme}

The full magnetohydrodynamic problem additionally requires a computation of 
the temporal evolution of the flow field. In this case, the
Navier-Stokes equation needs to be solved as well, which for a
precessing flow in the rotating frame of reference takes the form 
\begin{equation}
  \frac{\partial\vec{u}}{\partial t}  =
  -\vec{u}\cdot\nabla\vec{u}-{\nabla p} 
  +\frac{1}{\rm{Re}}\Delta\vec{u}
  -2({\rm{Po}\vec{k}}_{\rm{p}}\times\hat{\vec{z}}) +(\nabla\times\vec{B})\times\vec{B}+{\rm{Po}}\sin\alpha r \cos(\varphi + t)\hat{\vec{z}}.
\label{eq::navier}
\end{equation}
In addition to the terms commonly encountered in regular flow problems
describing nonlinear interactions and dissipation, we have three
additional contributions: the Coriolis force due to rotation, the
Lorentz force due to the interaction with the magnetic field, and the
Poincaré force accounting for acceleration due to perpetual change of
the orientation of the cylinder's axis of rotation. A detailed
derivation of the Navier-Stokes equation for a precessing system is
given e.g. in \citet{tilgner2005}.  

We conduct direct numerical simulations (DNS) of the coupled system of equations~(\ref{eq::induction}) and (\ref{eq::navier}) 
using the code {\tt{SpecDyn}}\cite{wilbert2022,wilbert_phd}, which is
based on a pseudo-spectral Fourier approach. The drawback of this
approach is the requirement of periodic boundary conditions in all
three directions. Since this rules out an application of a closed
cylindrical geometry, the scheme is complemented by a direct forcing
immersed boundary method (IBM) which allows the consideration of
nearly arbitrary shaped boundaries that prescribe the confinement for
the fluid flow when the boundary conditions on these boundaries are of
no-slip type. 
The fulfilment of the divergence-free condition for the velocity field
is ensured using a Pressure-Corrector method.
%where the
%pressure serves as a Lagrangian multiplier so that $\nabla \cdot
%\vec{u}=0$ is ensured. 
%(i.e. first, solve Equation~(\ref{eq::navier})
%without pressure, then project the intermediate solution on the
%divergence-free space by calculating the necessary pressure
%field). 
The same scheme is also applied to the magnetic field to
ensure that the solution of the induction equation is divergence-free
as required by Maxwell's equations. 
The code is parallelized using a two-dimensional processor grid. To apply the Fast Fourier Transform (FFT) to the distributed data along all spatial directions, the processor grid must be transposed, which is achieved by leveraging advanced MPI functions to perform an all-to-all communication of the processor data.
Tests have been conducted with various configurations that proved
accurate solutions with excellent convergence properties and nearly
perfect strong scaling properties. A detailed description is given in
\citet{wilbert2022} and in \citet{wilbert_phd} and a comparable
approach was used for numerical models of the VKS
dynamo\cite{kreuzahler2014,kreuzahler2017}. 

\subsection{Model set-up}

We consider a setup comparable to the kinematic models with a stagnant
outer layer, i.e. a setup given by a cylindrical fluid region with
radius $R=1$ and height $H=2$ embedded in a Cartesian domain with a
edge length $L=3$ so that 
the size of the box provides sufficient space to let the magnetic
field decline outside the cylindrical vessel. The cylindrical vessel
is implemented by imposing internal conditions on the fluid that
emulate the (virtual) container walls by enforcing no-slip conditions
at the surface of the container. 
Unlike in the kinematic (multi-)layer models we do not model any
special region representing the wall of the vessel because such a
layer would not be properly resolved.  
Instead we use a magnetic diffusivity $\eta=1$ for the fluid region,
whereas the 'non-liquid' region outside the vessel has a magnetic
diffusivity $\eta_{\rm{e}}=10$, which is supposed to emulate an
electrically isolating exterior.  
All simulations are carried out with a Reynolds number
${\rm{Re}}=\varOmega_{\rm{c}}R^2/\nu=6400$ which is the maximum
possible at the resolution of $N=256^3$ while still sufficiently
resolving the boundary layers at the fluid-wall interface of the
vessel \cite{wilbert_phd}.  

\subsection{Results}

Similar to the procedure in the kinematic case, we vary the
Poincar{\'e} number ${\rm{Po}}$, whereby different magnetic Reynolds
numbers are considered, beginning with high values. If a positive mean
growth rate is found for a particular set of parameters, further runs
are executed consecutively lowering the value of ${\rm{Rm}}$ until no
more dynamo action can be observed.  
Similar to the kinematic models we find  
a limited range for the preferred occurrence of a dynamo in the
interval $0.105\lesssim {\rm{Po}} \lesssim 0.125$. However, the
critical magnetic Reynolds number is significantly higher with a
minimum value of ${\rm{Rm}}^{\rm{crit}} \approx 5600$ at ${\rm{Po}} =
0.11$. The slightly larger value for the optimum ${\rm{Po}}$ compared
to the kinematic models can be explained by the (slightly) smaller
${\rm{Re}}$ (see also \citet{giesecke2024}).   
In the following we focus on the particular case ${\rm{Po}}=0.11$ and
${\rm{Rm}}=6500$ and compare the properties of this DNS solution with
the results found in the kinematic case. The chosen paradigmatic
example well characterizes the DNS solutions found very close to the
threshold while allowing a faster growth with shorter time period
until a statistically steady state is reached. 

\subsubsection{Temporal evolution of the kinetic energy}

\begin{figure}[b!]
\begin{center}
\subfloat[][]{\includegraphics[width=0.49\textwidth]{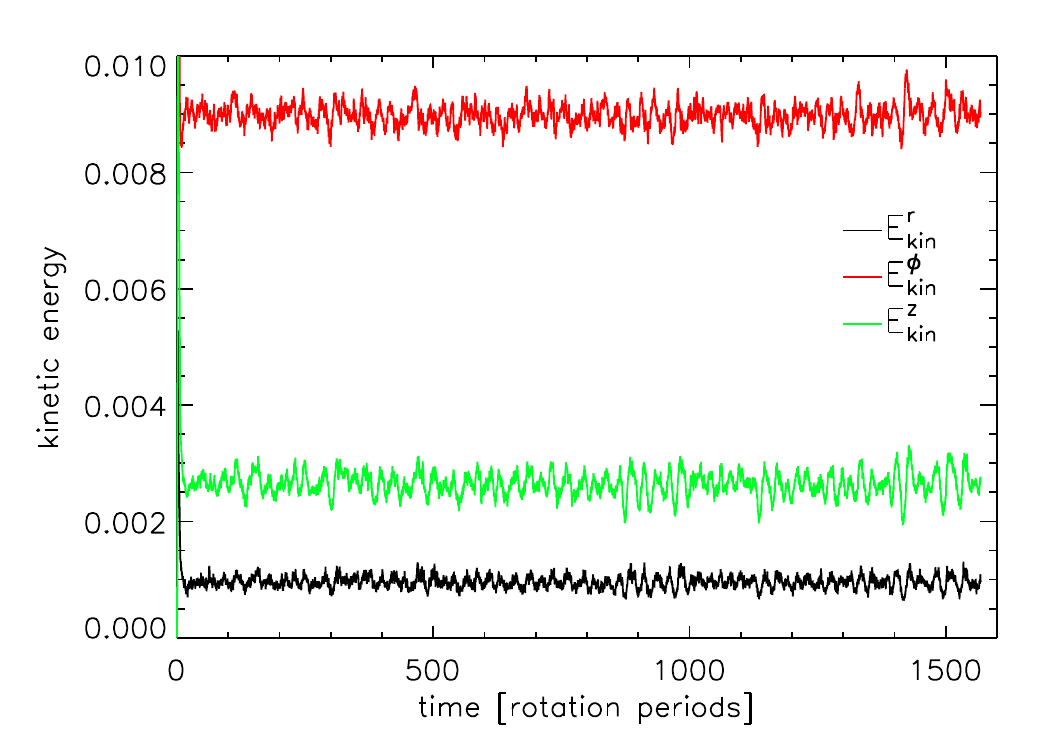}\label{fig::kinener_vs_tim_DNS}}
\subfloat[][]{\includegraphics[width=0.49\textwidth]{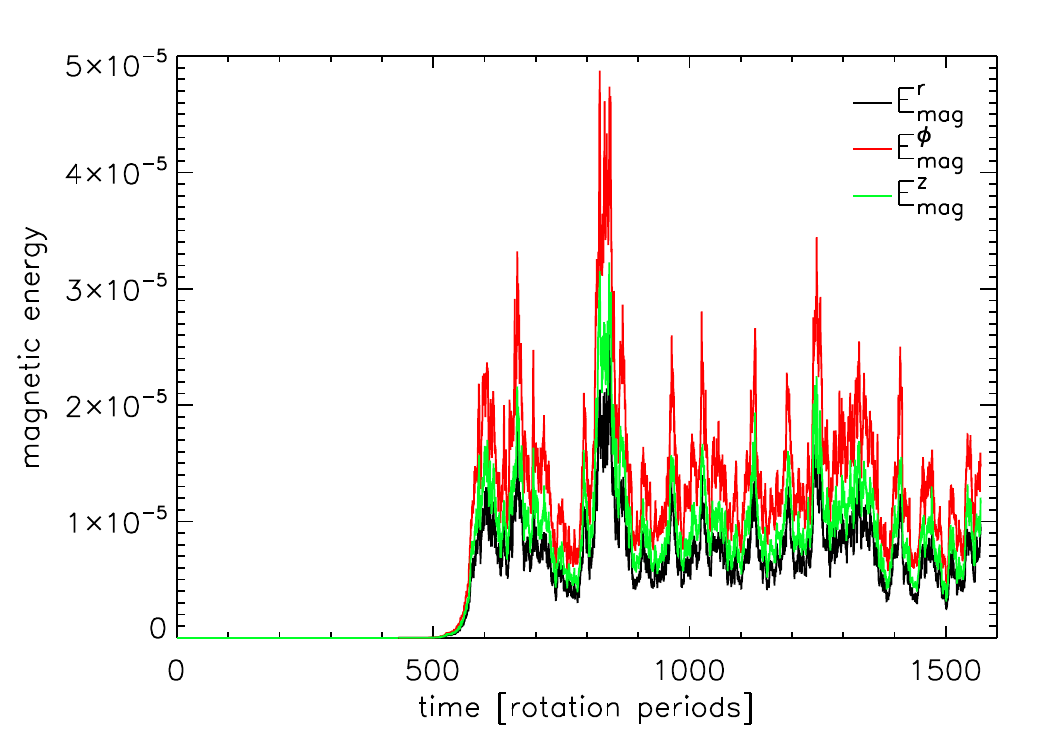}\label{fig::magener_vs_tim_DNS}}
\end{center}
\caption{Temporal evolutions of the kinetic (a) and magnetic (b) energy for ${\rm{Re}}=6400, {\rm{Rm}} = 6500$ and ${\rm{Po}} = 0.11.$}
\label{ek-em}
\end{figure}

The temporal evolutions of kinetic and magnetic energy $E_{\rm{k}}$
and $E_{\rm{m}}$ are displayed in Fig.~\ref{ek-em}, where $E_{\rm{k}}$
and $E_{\rm{m}}$ are defined as  
\begin{equation}
    E_{\rm{k}}  =  \frac{1}{2L^3}\int\limits_{L^3}\left|\vec{U}(\vec{r},t)\right|^2 d V,
    \mbox{ and }
    E_{\rm{m}}(t) =  \frac{1}{2L^3}\int\limits_{L^3}\left|\vec{B}(\vec{r},t)\right|^2 d V
\end{equation}
and where $L$ denotes the edge length of the cubic domain.
The simulations start with the pure hydrodynamic case, which is
processed until a quasi-stationary statistically steady state is
reached. A detailed description of the hydrodynamic behavior is given
in \citet{wilbert2022}. Here we focus on the properties of the
magnetic field in the dynamo case and the resulting feedback on the
velocity field. After switching on the magnetic induction at $t=300$,
the magnetic energy begins to grow for about $100$ cylinder rotations
and finally goes into saturation. 
Both, kinetic and magnetic energy are dominated by contributions of
the azimuthal components of the related field. However, the dominance
for the magnetic field is not particularly pronounced and the induced
magnetic energy $E_{\rm{m}}$ remains much smaller than $E_{\rm{k}}$. 
We find that, even in the saturated state, the presence of the
magnetic field  does not substantially alter the structure and the
amplitude of the fluid flow via the Lorentz force.  
A striking property of the magnetic energy is that, unlike the kinetic
energy, which remains relatively steady, the magnetic energy is
characterized by strong occasional peaks where $E_{\rm{m}}$ suddenly
rises $3$ to $5$ times above its average value. 
There are no related substantial changes in the flow field due to the
back-reaction of the magnetic field so that the bursts can be clearly
assigned to corresponding peaks in the interaction parameter $\zeta$,
which compares the amplitude of the Lorentz force to the inertial term
in the Navier-Stokes equation: 
\begin{equation}
\zeta=\displaystyle\frac{\left|(\nabla\times\vec{B})\times\vec{B}\right|}{\left|\mu_0(\vec{u}\cdot\nabla)\vec{u}\right|}.
\end{equation}
\begin{figure}
    \includegraphics[width=0.75\textwidth]{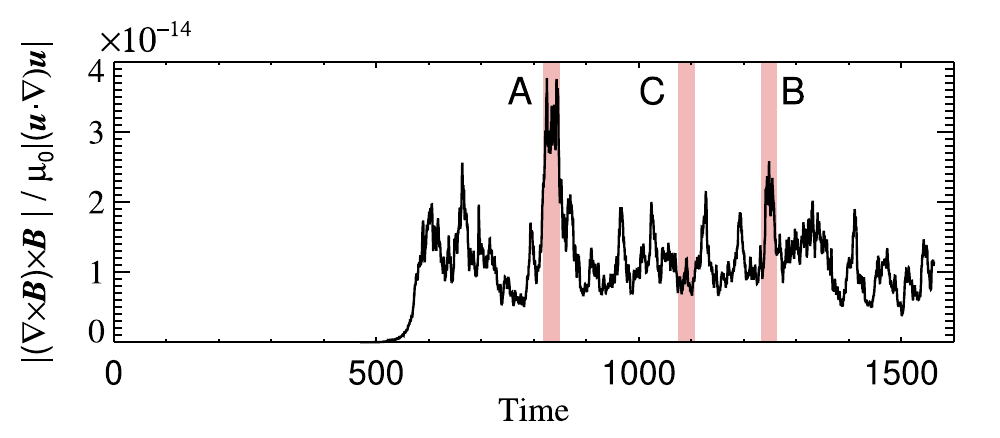}
    \caption{\label{fig::interaction}Interaction parameter showing the
      bursts of magnetic energy in comparison with the kinetic
      energy. The vertical lines mark the time steps as visualized in
      Figure~\ref{fig::snapshots} and the shaded area shows the period
      for which animations are available in the supplementary data of
      this study \cite{supplement2}.} 
\end{figure}
For a self-excited dynamo $\zeta$ follows $\sim {B^2}/{U^2}$ (see
\citet{miralles2015}), which is indeed the case as shown in
Figure~\ref{fig::interaction}, 
where we see similar irregular peaks due to the bursts shown by the
magnetic energy in Figure~\ref{fig::magener_vs_tim_DNS}. 
However, the impact of the magnetic field remains small even in the
case of a burst.  

\begin{figure}
    \subfloat[][]{\includegraphics[width=0.245\textwidth]{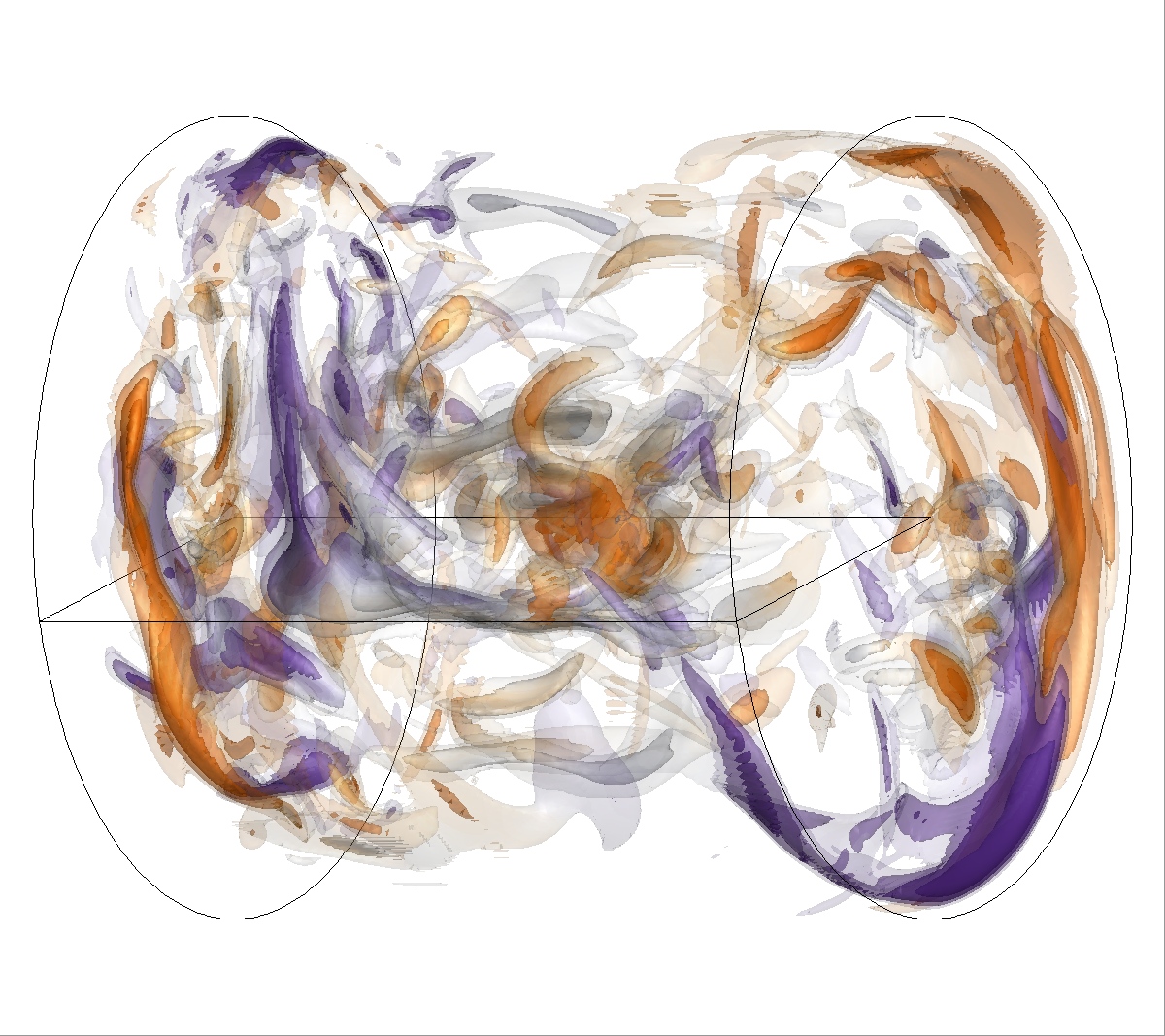}\label{fig::snapshot_a1}}
    \subfloat[][]{\includegraphics[width=0.245\textwidth]{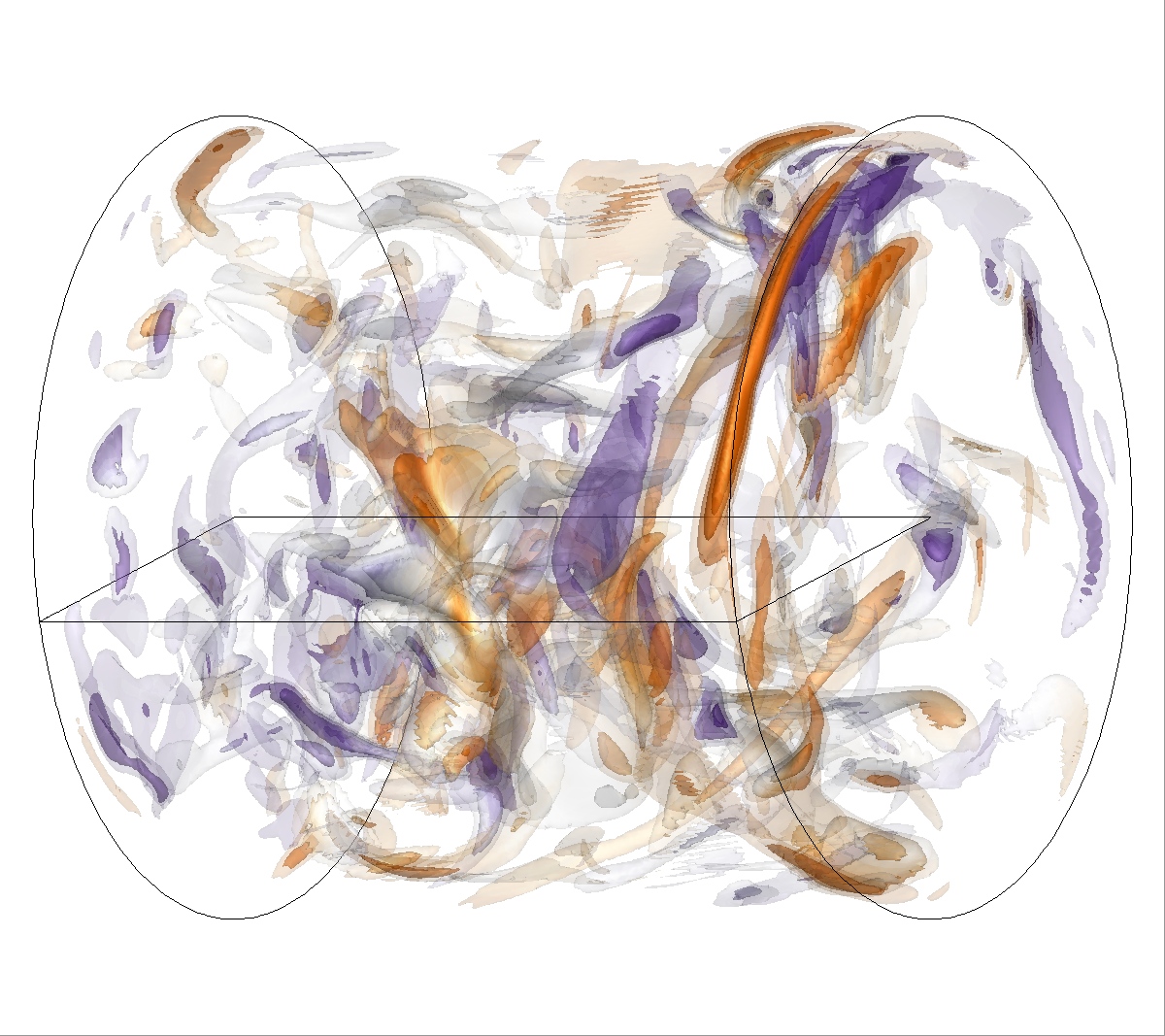}\label{fig::snapshot_a2}}
    \subfloat[][]{\includegraphics[width=0.245\textwidth]{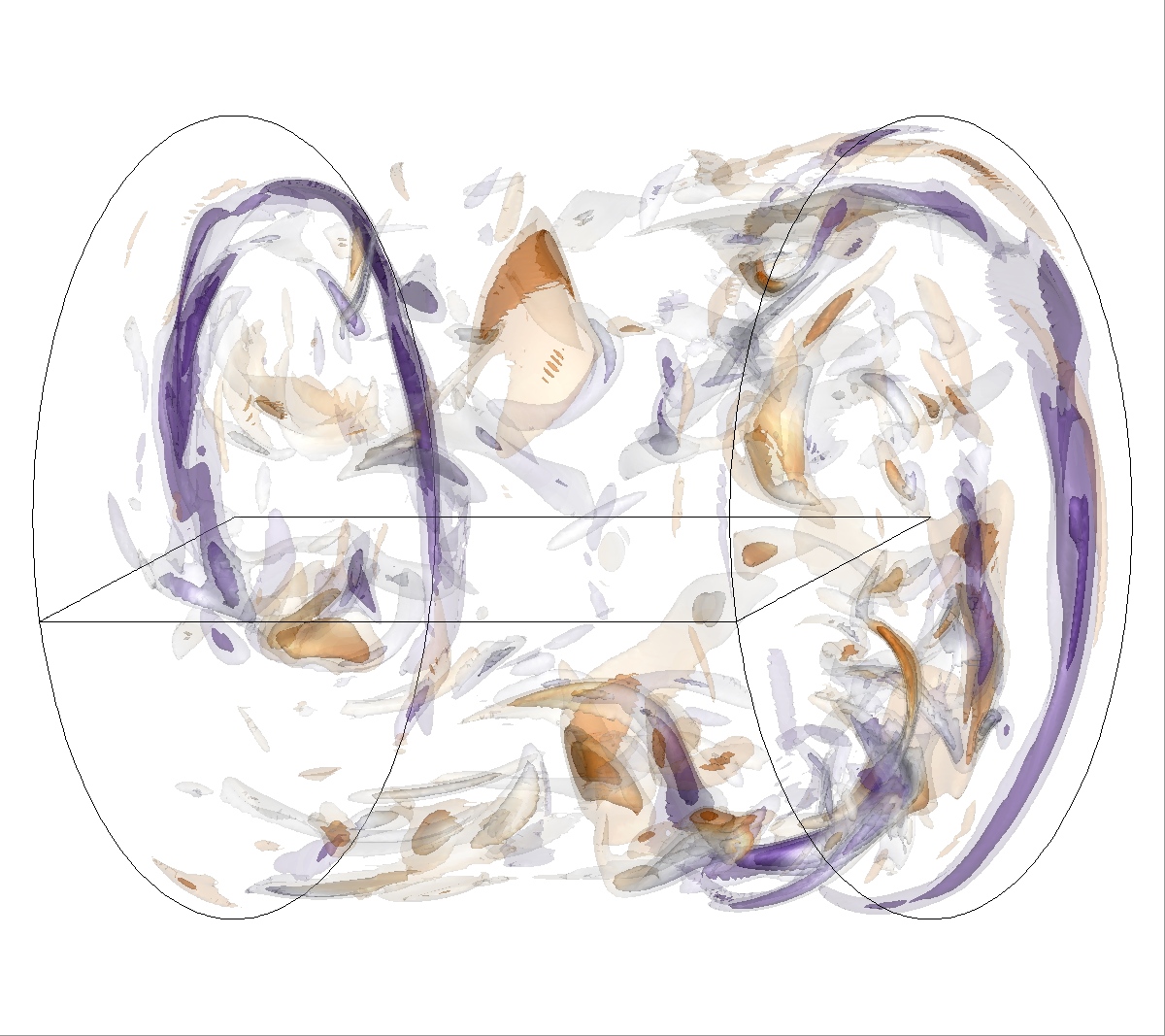}\label{fig::snapshot_a3}}
    \subfloat[][]{\includegraphics[width=0.245\textwidth]{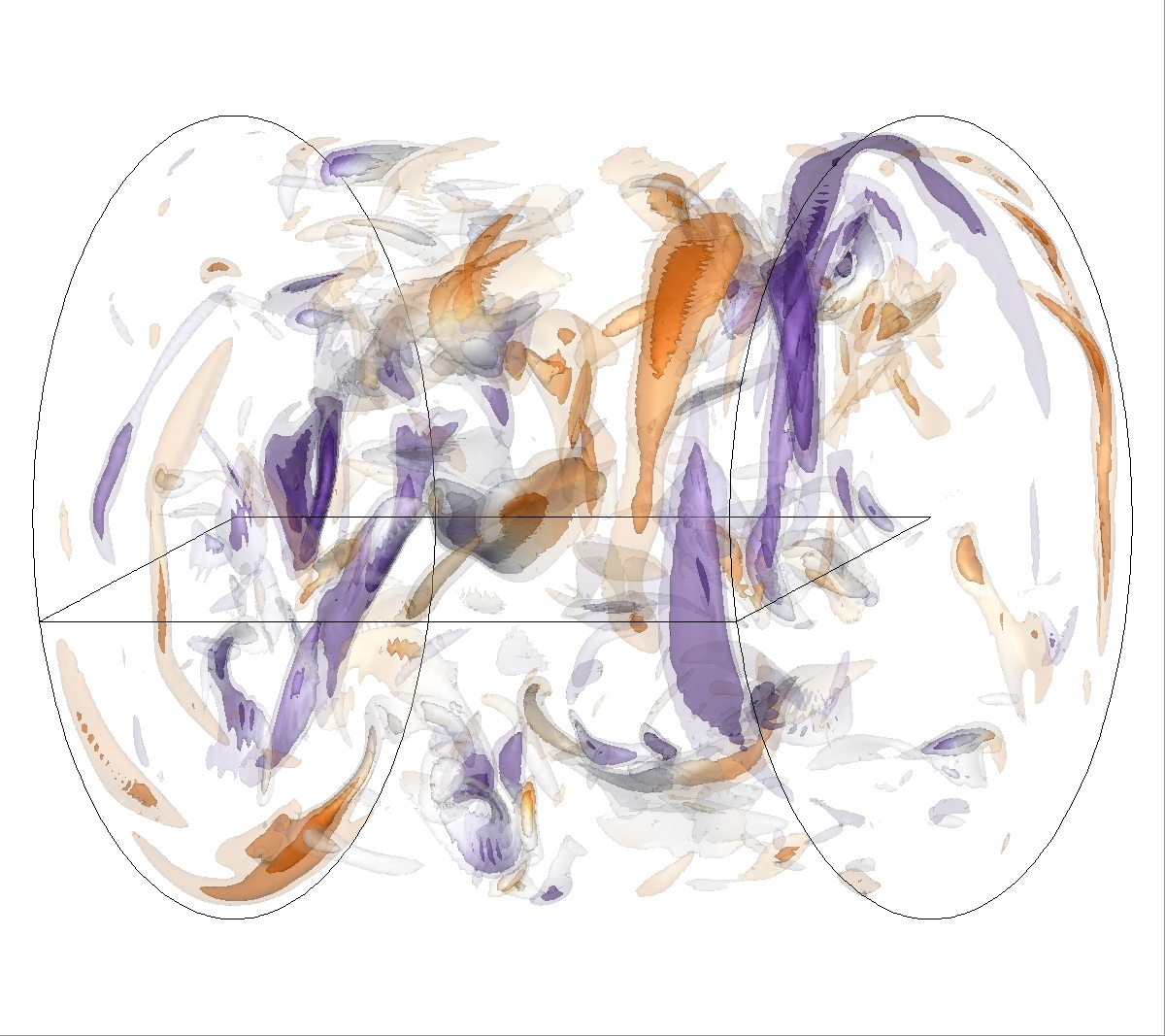}\label{fig::snapshot_a4}}
\\[-1cm]
    \subfloat[][]{\includegraphics[width=0.245\textwidth]{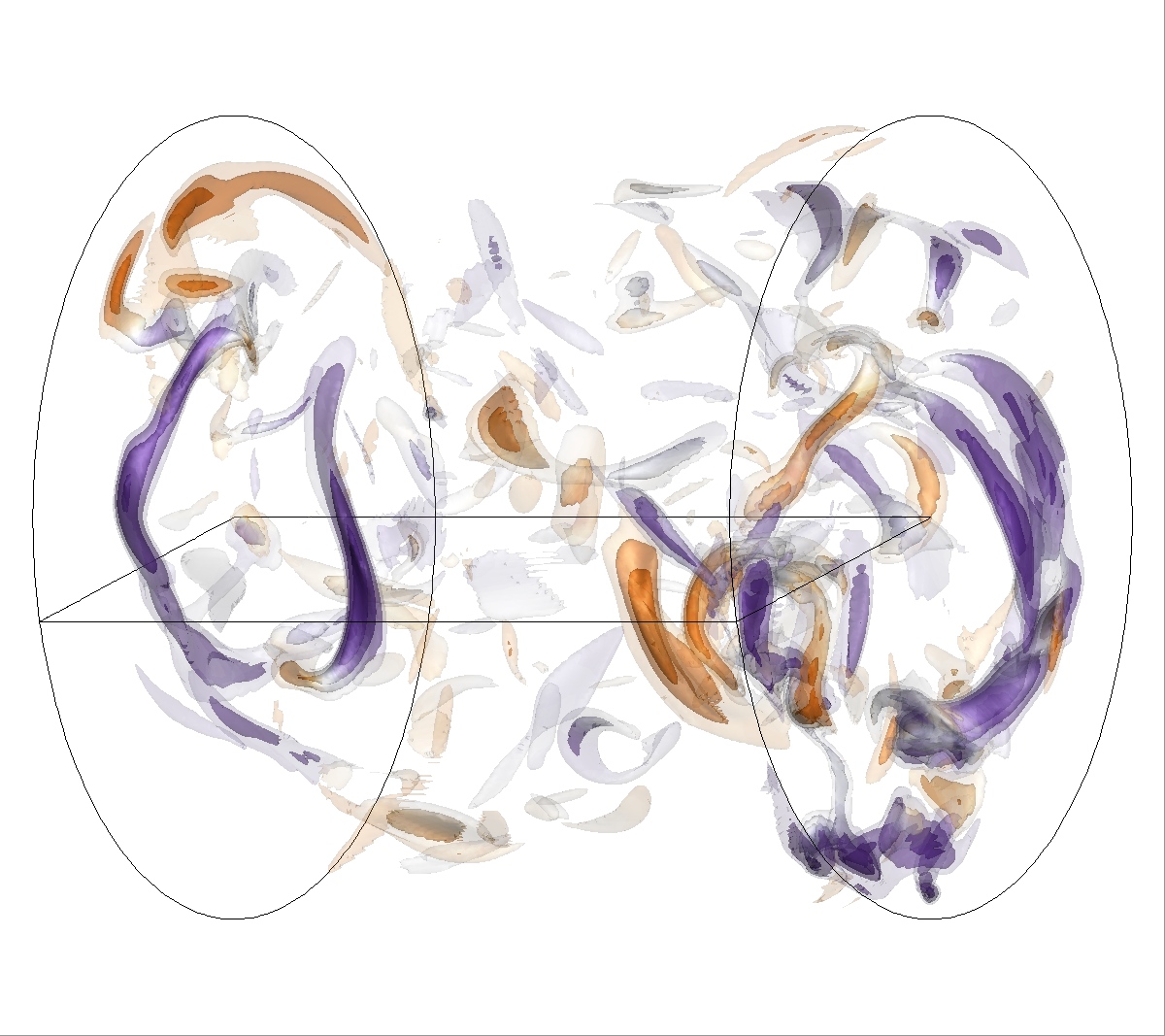}\label{fig::snapshot_b1}}
    \subfloat[][]{\includegraphics[width=0.245\textwidth]{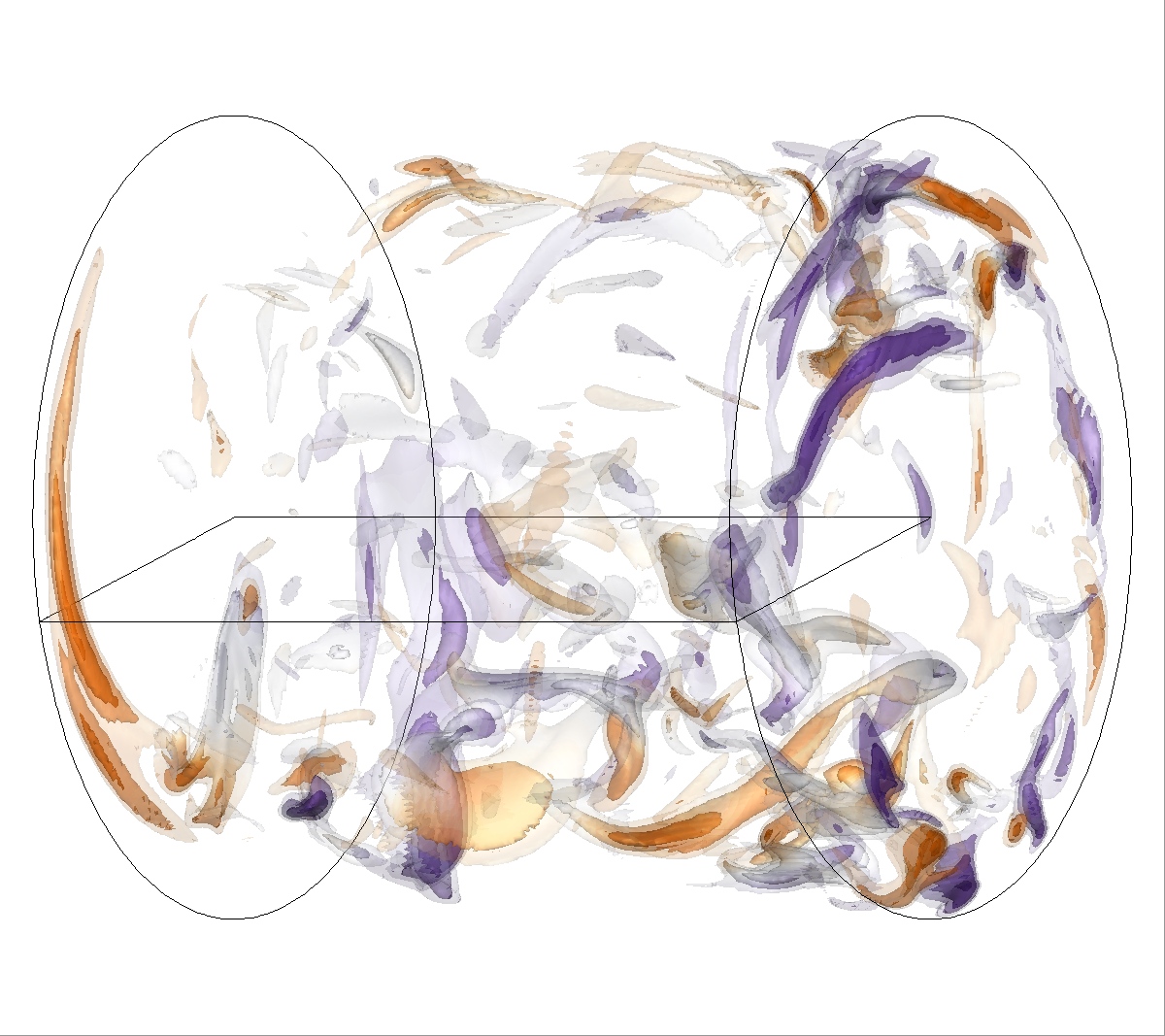}\label{fig::snapshot_b2}}
    \subfloat[][]{\includegraphics[width=0.245\textwidth]{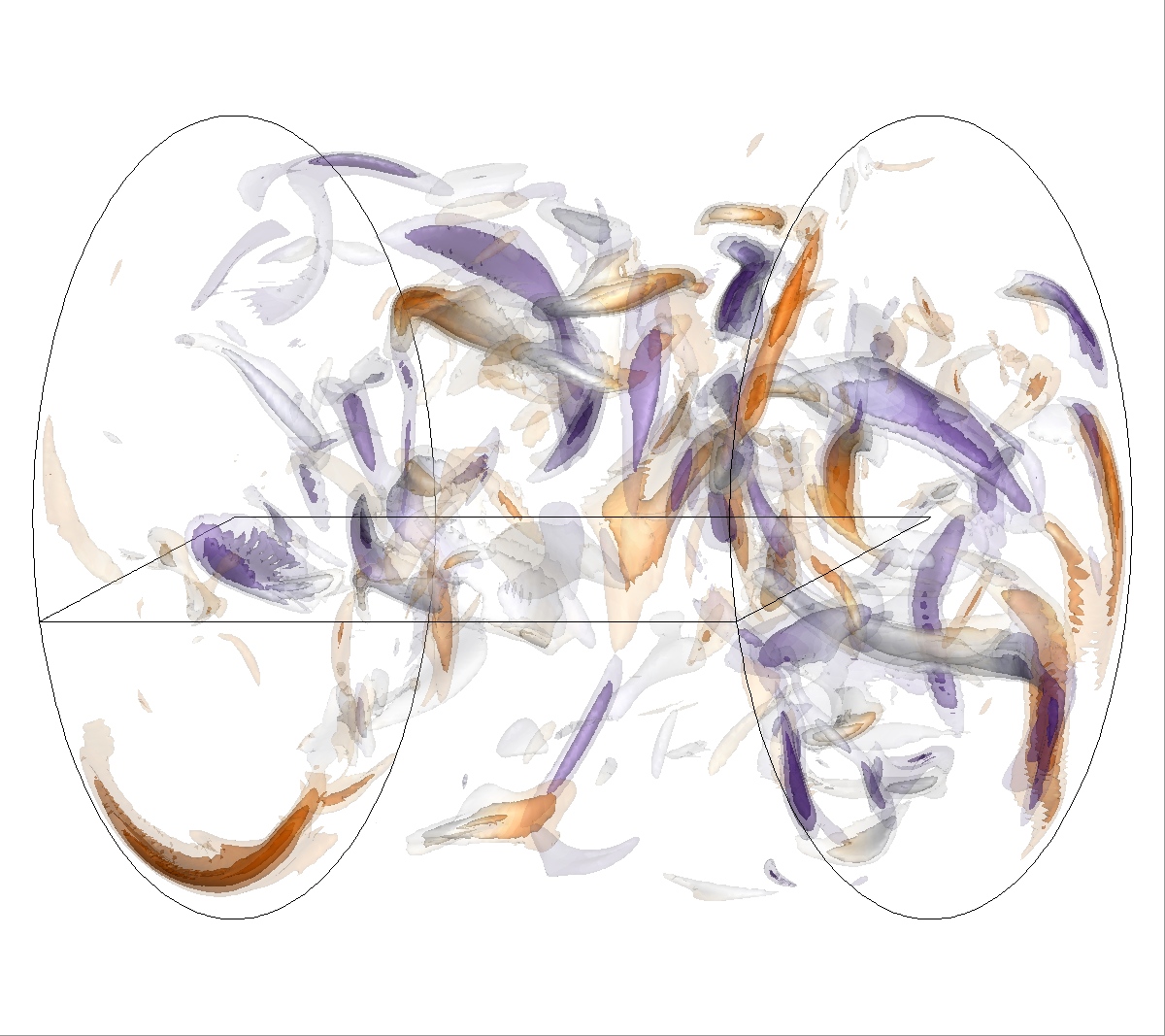}\label{fig::snapshot_b3}}
    \subfloat[][]{\includegraphics[width=0.245\textwidth]{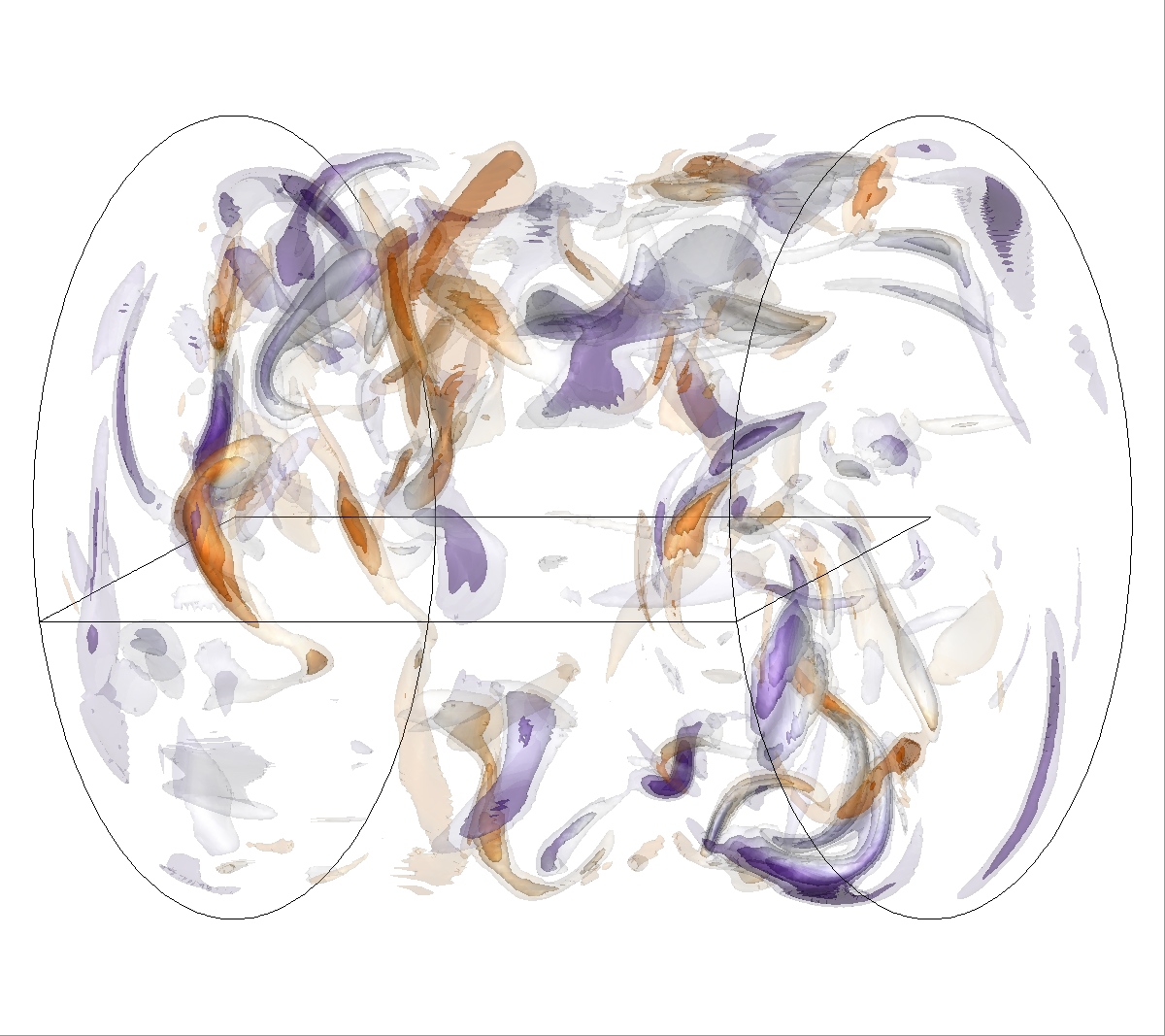}\label{fig::snapshot_b4}}
    \\[-1cm]
    \subfloat[][]{\includegraphics[width=0.245\textwidth]{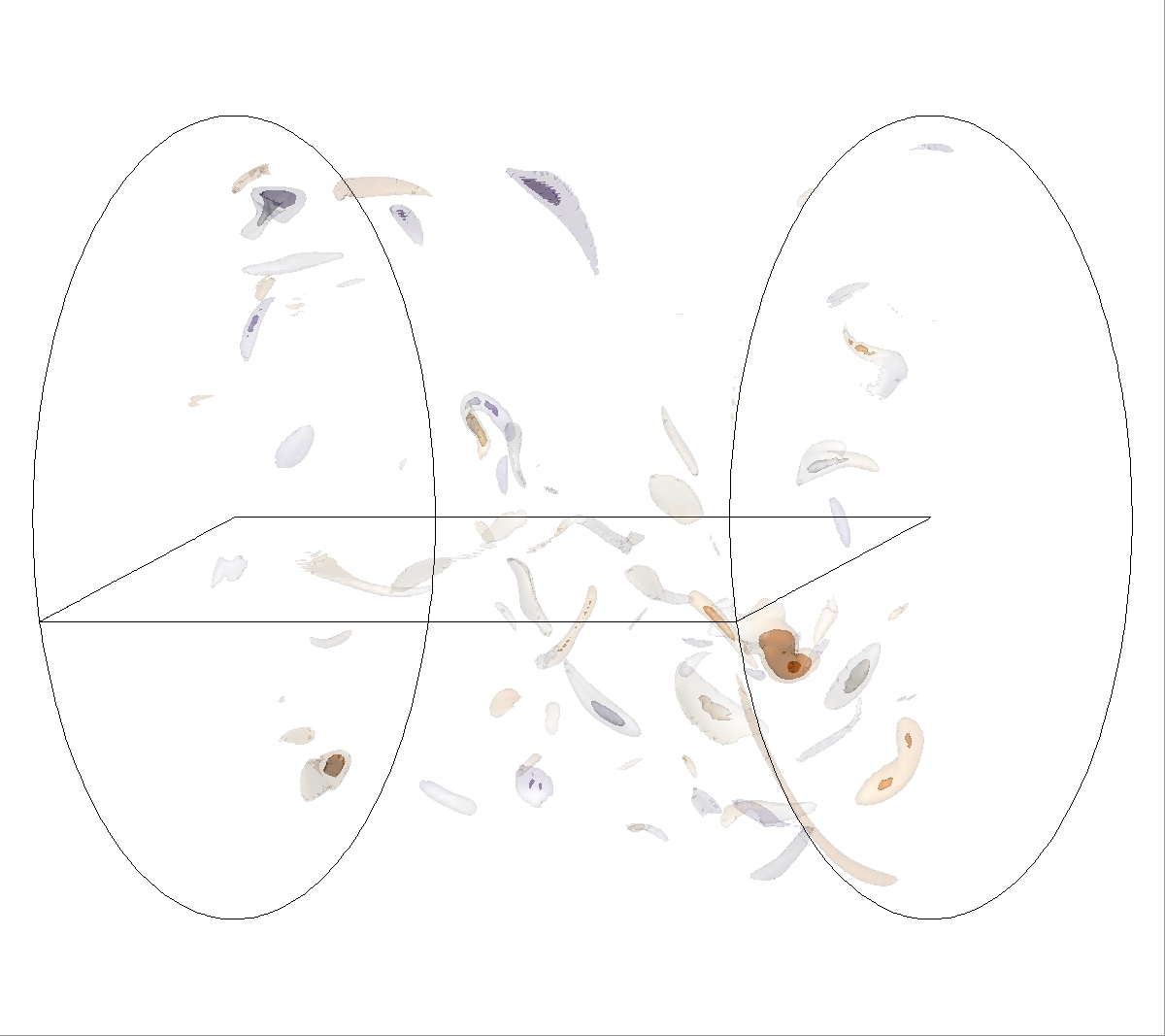}\label{fig::snapshot_c1}}
    \subfloat[][]{\includegraphics[width=0.245\textwidth]{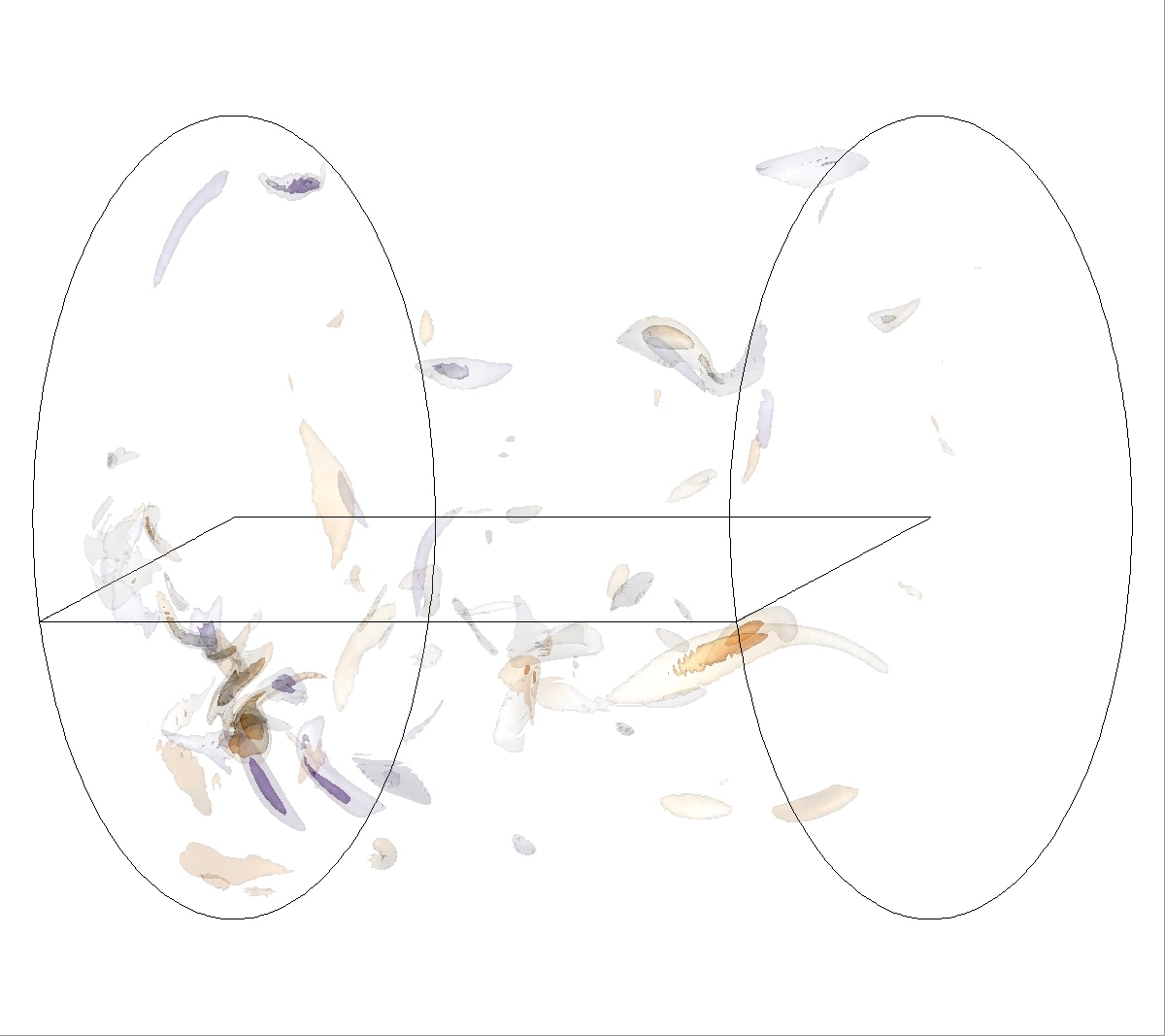}\label{fig::snapshot_c2}}
    \subfloat[][]{\includegraphics[width=0.245\textwidth]{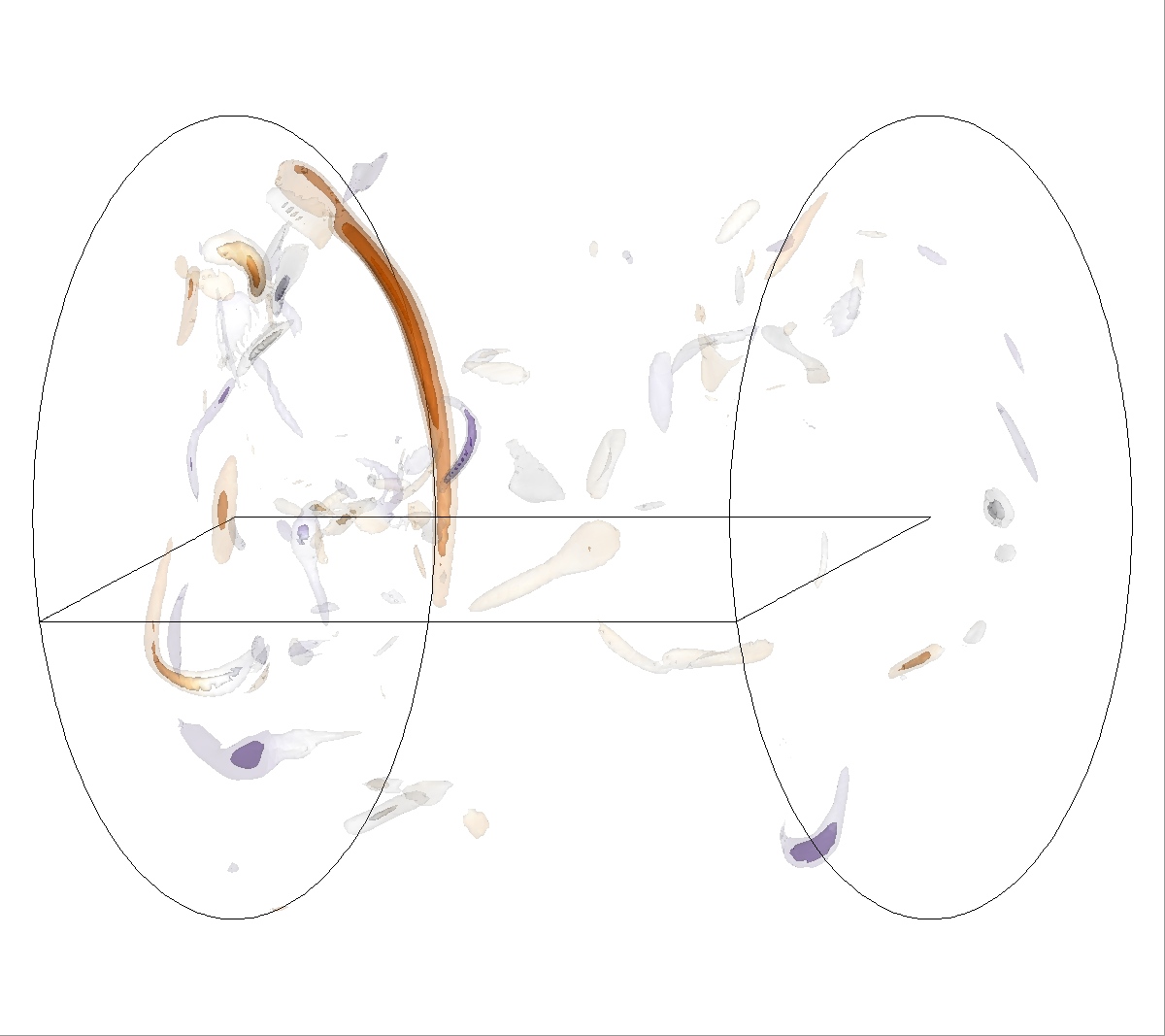}\label{fig::snapshot_c3}}
    \subfloat[][]{\includegraphics[width=0.245\textwidth]{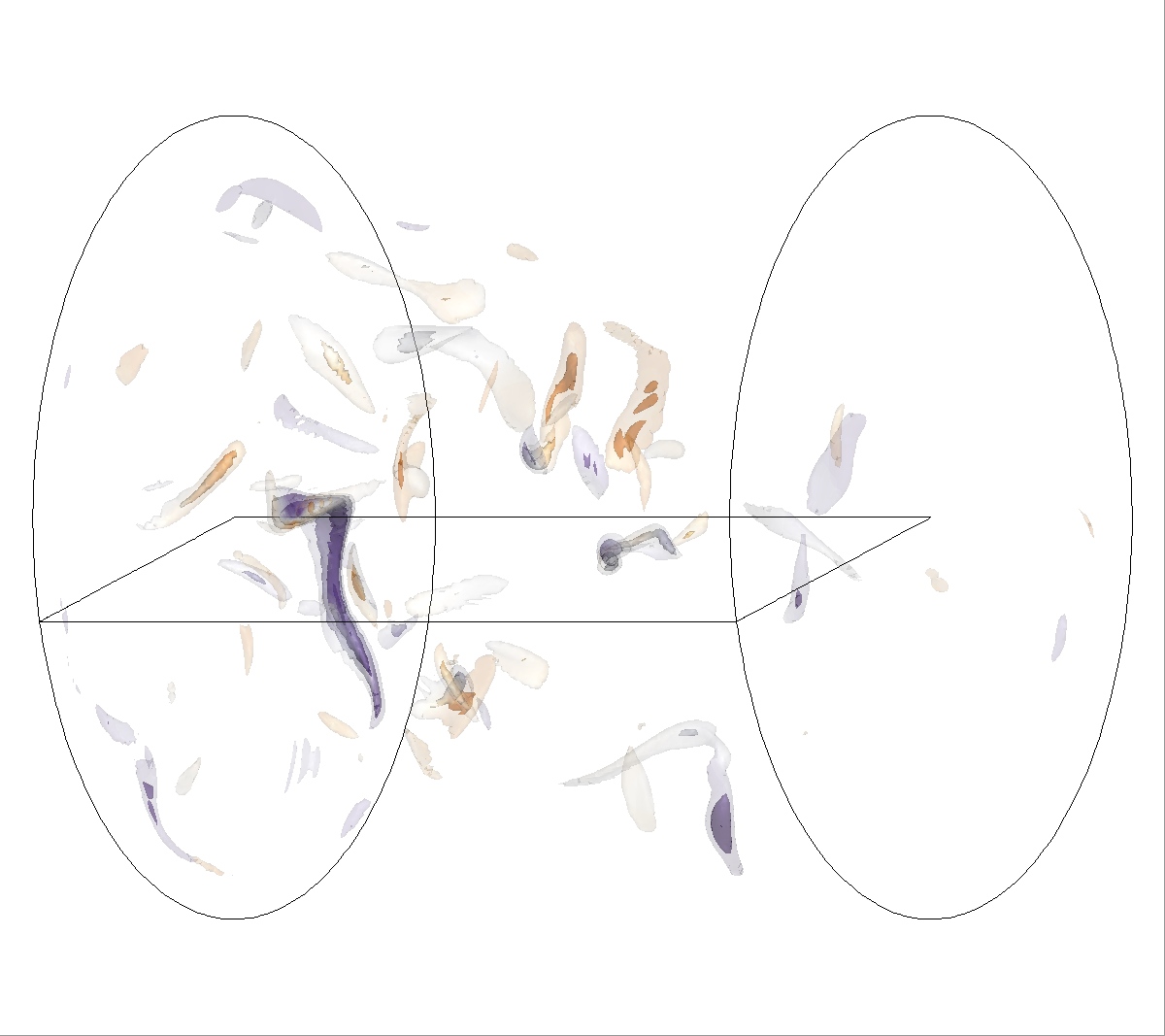}\label{fig::snapshot_c4}}
       \caption{\label{fig::snapshots}Snapshot of the magnetic energy
         at $5\%, 15\%, 30\%,$ and $50\%$ of the respective average
         value.  
    Each row presents four arbitrary snapshots taken within the
    related period $A, B$, and $C$ (from top to bottom).     
    The color codes represent the radial component of the magnetic
    field.
       }
\end{figure}

\begin{figure}
    \includegraphics[width=0.65\textwidth]{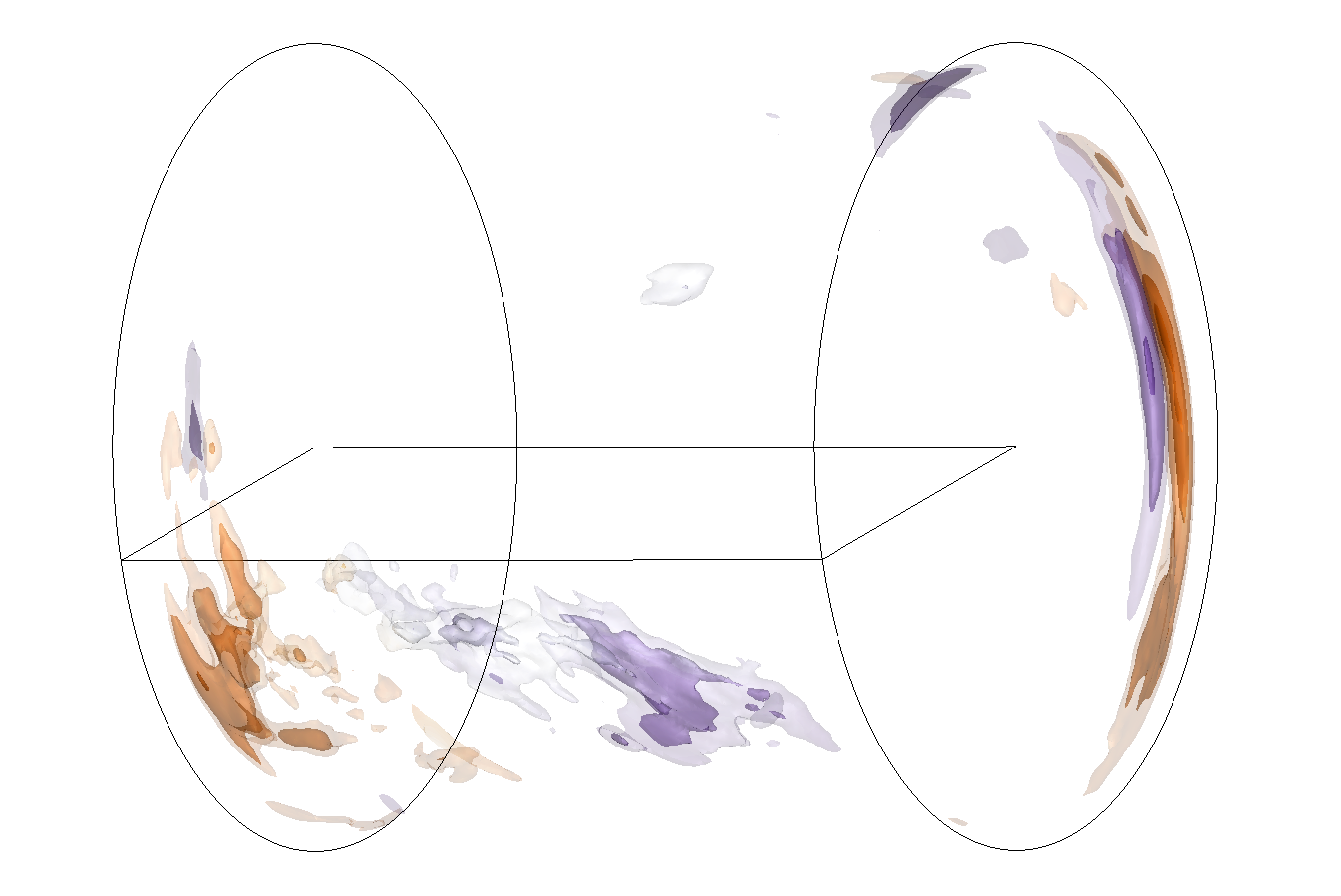}
    \caption{\label{fig::mean_magnetic_field}Time averaged iso-surfaces of the magnetic energy. The colors denote the radial magnetic field component.}
\end{figure}

In the following, we discuss three characteristic periods as labeled
by capital letters A, B, (two paradigmatic peaks) and C (quiescent
period) in Figure~\ref{fig::interaction}. 
Figure~\ref{fig::snapshots} shows three rows each with four snapshots
of the spatial distribution of the magnetic energy taken at four
arbitrary times within each related period. 
Animated sequences showing the time development within the marked
periods can be found in the supplementary material of this
study\cite{supplement2}. 
We recognize a small scale field  formed by flux patches
distributed randomly in space and time. 
These patches are short-lived despite the strong concentration of
magnetic energy, and even if they appear regularly, no corresponding
period can be assigned. 
The magnetic field patches are slightly elongated in the azimuthal
direction and 
during a typical peak period, we see an increased excitation of
medium-scale concentrations of magnetic energy in terms of elongated
arc-like magnetic structures near the end caps
(Figure~\ref{fig::snapshot_a1}--\ref{fig::snapshot_b4}). However,
these mid-scale magnetic field structures remain isolated and decay
within 1-2 rotation periods, whereby the resulting small scale patches
have a rather isotropic character and are distributed within the
entire volume of the cylinder. 
While the distribution of the magnetic energy in individual snapshots
appears to be largely chaotic (with the exception of the bursts), this
is no longer the case when we look at the time average in the
turntable system. Applying a long averaging time of more than 200
rotation periods, we obtain a rather regular distribution of magnetic energy as shown in Figure~\ref{fig::mean_magnetic_field}. Similar to the low-efficient branch of the kinematic models we find elongated structures close to the end
caps, which are a result of the statistical distribution of the magnetic field energy. 
Furthermore, and in contrast to the kinematic models, we
recognize different symmetries with respect to the mid-plane of the
cylinder, which are expressed in the different signs for the
time-averaged radial field. We therefore see both, a
dipole-like state and a quadrupole-like state, which are
alternately dominant at irregular intervals.
Due to the small-scale structure, however, the characterization as a dipole or quadrupole field is not really justified. 
In this sense, the results are consistent with other
dynamo simulations in which a mechanical forcing was utilized as
driving mechanism (see e.g. \citet{lin2016}, \citet{cebron2019},
\citet{reddy2018}).
In agreement with Landeau's conclusions\cite{landeau2022}, we therefore confirm that it
remains unclear whether mechanical forcing can generate a large-scale
magnetic field, and apart from experimental investigation, only
simulations with massively reduced viscosity (i.e. massively larger
Reynolds number) can help to clarify this question, which is currently
prevented by insufficient performance of numerical algorithms and HPC systems.

\subsubsection{Back-reaction and kinetic power spectrum}

\begin{figure}
%    {\includegraphics[width=0.65\textwidth]{velocity_peak_a_step13200.pdf}}%\label{fig::snapshot_d1}}
    {\includegraphics[width=0.65\textwidth]{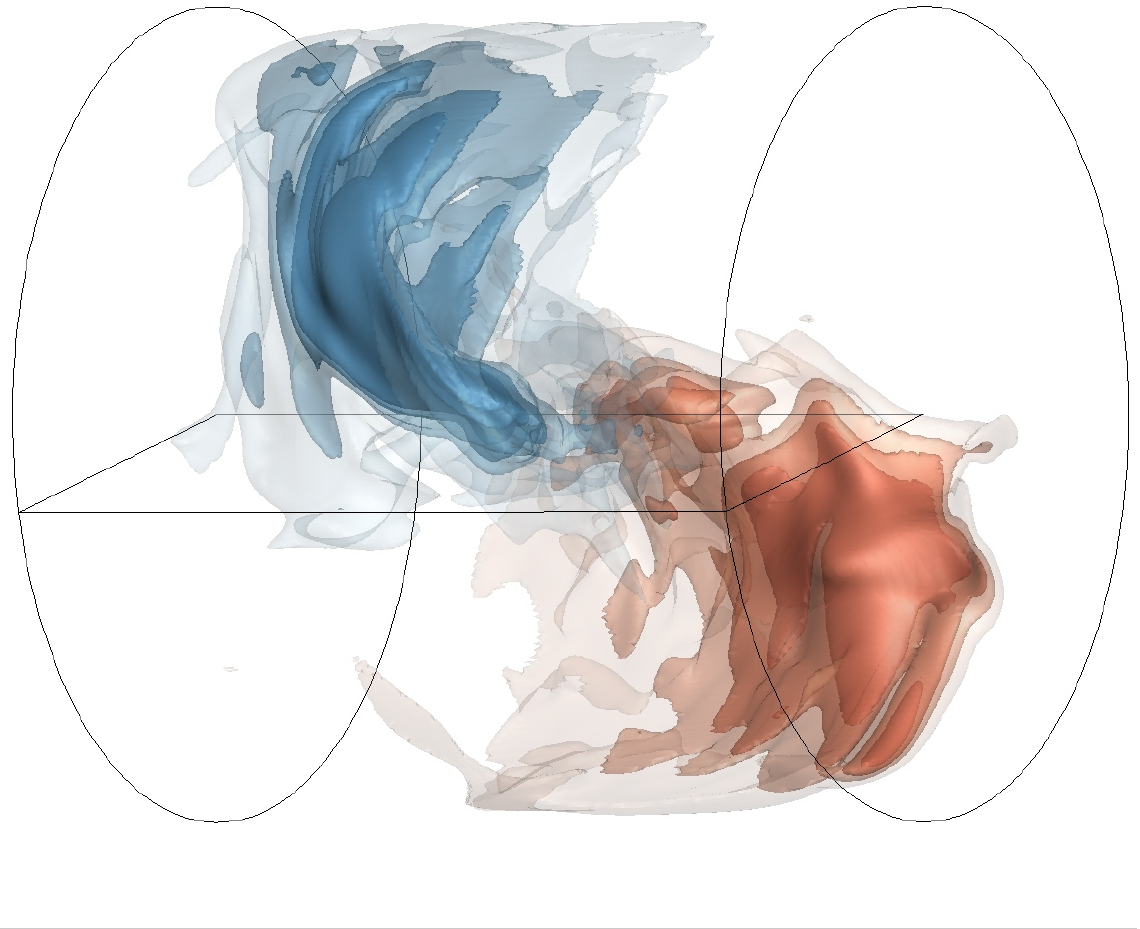}}%\label{fig::snapshot_d1}}
%    \subfloat[][]{\includegraphics[width=0.485\textwidth]{vorticity_13200.jpg}\label{fig::snapshot_h1}}
    \caption{\label{fig::snapshot_velocity}Snapshot of the axial velocity 
    at an arbitary timestep during the first peak A. 
    The velocity field 
    looks rather similar independently of being at a peak state or in
    the quiescent region. An animations of the axial component shown
    in this plot can be found in the supplementary material.
    }
\end{figure}

In contrast to the magnetic field, the velocity field is always
dominated by large scales 
as shown in Figure~\ref{fig::snapshot_velocity} by means of a
paradigmatic time-snapshot of the three-dimensional distribution of
the axial velocity component $u_z$. 
This figure shows the typical structure of the velocity, which looks
very similar at any time, regardless of whether with or without a
magnetic field, or whether within a peak period or within a
magnetically quiet section. 
%In particular, we see no obvious change in the velocity field due to
%the magnetic field when regarding the kinetic energy or the dynamical
%structure.  
%
Indeed, a reduction of the speed amplitude and/or of the field structure is not
absolutely necessary, because here we deal with a small-scale
dynamo for which theoretical considerations (in a mean-field model) have shown that the feedback via the Lorentz force can be described by 
an increased diffusivity or hyperdiffusivity $\propto \nabla^4\vec{B}$
(\citet{subramanian2003}). 
\begin{figure}[b!]
    \includegraphics[width=0.48\textwidth]{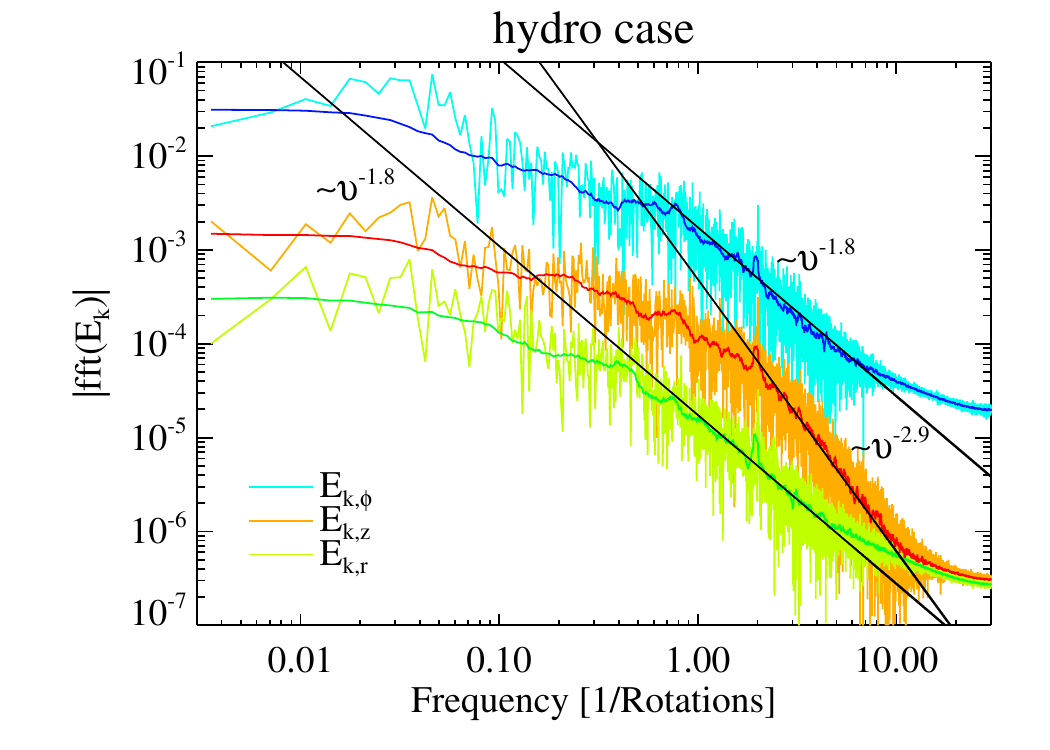}
    \hspace*{-0.4cm}
    \includegraphics[width=0.48\textwidth]{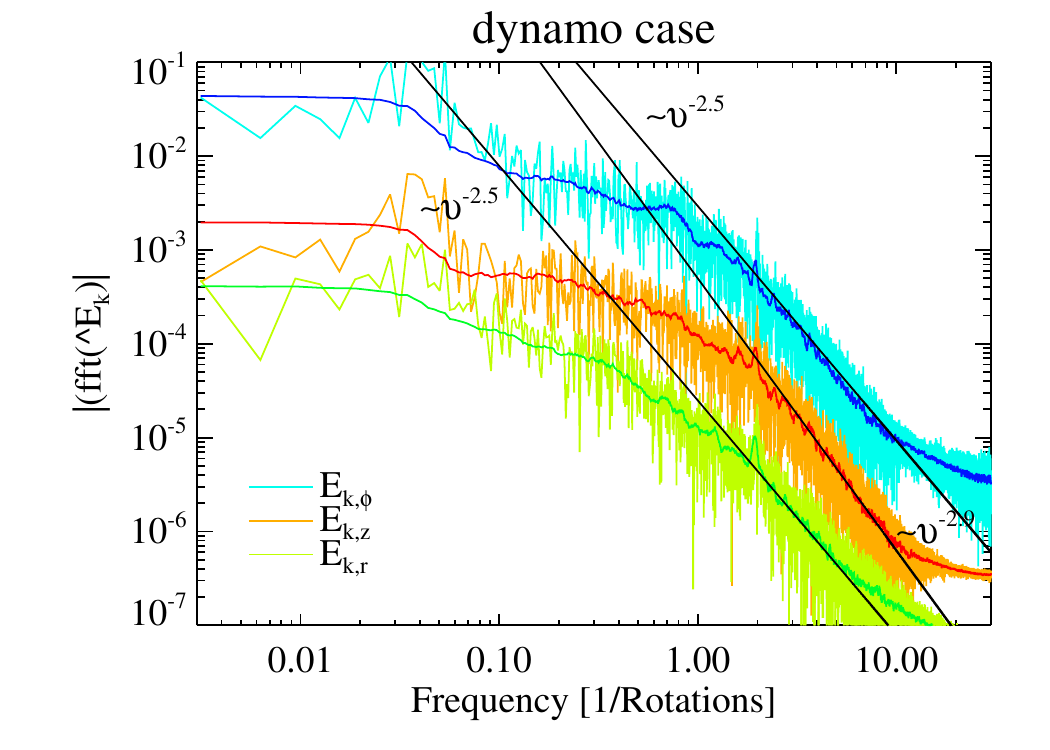}
    \caption{\label{fig::tspec_ekin}Temporal power spectrum of the velocity field in the hydro case (a) and in the dynamo case (b).}
\end{figure}
A response of the fluid flow can be seen in a subtle way when
looking at the temporal power-spectra of kinetic
energy. Figure~\ref{fig::tspec_ekin} compares the temporal power
spectrum of the kinetic energy for the hydrodynamic case (left panel)
with the spectrum for the dynamo case (right panel). Here we assume
that in the high frequency range the kinetic energy of the individual
components $r, \varphi$ and $z$ scales according ${\hat{E}}_{\rm{k}} \propto \nu^{\alpha}$ where ${\hat{E}}_{\rm{k}}$ is the energy from the Fourier
component with the frequency $\nu$ and $\alpha$ is a spectral index
that turns out to be different for the horizontal coordinates  
($r,\varphi \rightarrow \alpha_{\perp}$) and the vertical coordinate
($z\rightarrow \alpha_{\parallel}$). 
In the hydrodynamic case, the contributions to high frequencies by the axial 
component $z$ (spectral index $\alpha_{\parallel}=-2.9$) drop
significantly faster than the contributions of the horizontal
components $r$ and $\varphi$ (spectral index $\alpha_{\perp}\approx
-1.8$).  
In the dynamo case, the spectral index of the horizontal components
approaches $\alpha_{\perp}\approx -2.5$ which is close to the axial
component, which practically does not change compared to the
hydrodynamic case.  
In summary, in the magnetic case we find an isotropization of the spectrum and a damping of the temporal fluctuations, which is expressed in the steeper spectrum.

%%%%%%%%%%%%%%%%%%%%%%%%%%%%%%%%%%%%%%%%%%%%%%%%%%%%%%%%%%%%%%%%%%%%%%%%%%%%%%%%%%%%

\section{Summary and Conclusions\label{sec::05_conclusions}}

We have run numerical simulations of precession driven dynamo action
in a cylindrical cavity. We examined simplified kinematic models as
well as a particular setup of a similar system solved by a
self-consistent three-dimensional magnetohydrodynamic approach. 
The kinematic models reveal the existence of two different branches
for dynamo action, the realization of which depends on the
configuration of the outer boundary regions. The first branch is
characterized by a highly localized and stationary dynamo that only
occurs at rather large magnetic Reynolds number. The second type of
dynamo is of oscillatory type with the magnetic field filling a larger
proportion of the fluid domain. This type can occur at much lower
magnetic Reynolds number, which, however requires suitable boundary
conditions in terms of either a thin outer layer with large magnetic
diffusivity or a comparable emulation with pseudo-vacuum BC. 

We have not investigated here at which exact parameters the transition
from one branch to the other takes place, since the experimental
realization must in any case start from a basically infinitely large
outer volume of an insulator, namely the air in the laboratory.  
The kinematic model with two outer layers (for the container walls and for the lab environment), which we assume best reflects the experimental
setup, has a critical magnetic Reynolds number of
${\rm{Rm}}_{\rm{crit}}\approx 1300$, which  
agrees surprisingly well with the value derived by
\citet{goepfert2016} obtained from simulations in a precessing cube. 
Our solutions with external outer walls and an increased diffusivity
(i.e. a decreased electrical conductivity) are consistent with previous studies\cite{avalos2003,avalos2005}, who found a
monotonic reduction of the dynamo threshold for increasing
conductivity of an outer wall. Furthermore, these studies also reported a reduction
for oscillatory solutions with an optimum at a particular thickness of
the outer wall, which was explained by the increased dissipation
resulting from eddy currents induced in the outer wall.  

We also used direct numerical simulations in order to investigate the time evolution of the velocity field, the impact of the magnetic field through the Lorentz force and the consistency with the kinematic models. 
Unfortunately, it is not possible to achieve the large hydrodynamic Reynolds
number, as it is relevant in the experiment, due to the
limitations of the numerics and the hardware,  
so that we find dynamo solutions in the DNS only for relatively large
magnetic Reynolds numbers. 
This means that we effectively deal with a magnetic Prandtl number of
the order of one and that the magnetic field is of small-scale
type. At this stage it cannot be concluded with certainty that this
behavior will change if we move to larger (hydrodynamic) Reynolds
numbers, as there are currently insufficient parameter studies
available for such scaling.  
Nevertheless, for the parameters considered, a small-scale burst
dynamo is the typical solution. These bursts are characterized by
peaks in magnetic energy, which represent the sudden excitation of
large-scale magnetic field structures. Since these structures are
isolated and not sustained throughout the entire volume, they are
rapidly destroyed by  magnetic diffusion.  
Our results align with findings obtained for example by
\citet{etchevest2022, richardson2012, hughes2001}, which conclude that
increasing the magnetic Prandtl number leads to a transition from
large-scale dynamos to small-scale dynamos.  

The small scale characteristics of the magnetic fields from the DNS
are clearly different from  the properties of the field obtained in the kinematic
simulations. However, if the magnetic field from the DNS is averaged
over a sufficiently long period of time so that the occasional
occurrence of the strongly localized patches is also averaged out, the
structure is comparable to the structure of the weakly efficient
branch in the kinematic models, which only appears at large magnetic
Reynolds numbers. This explains, at least in part, the high magnetic
Reynolds number required to obtain a dynamo in the DNS. 

In order to obtain a dynamo solution that can also be realized in the
planned experiment, it would be necessary to find a path that leads to
the efficient dynamo branch with low ${\rm{Rm}}^{\rm{c}}$. Whether
this exists at all in the full MHD case, and how this can be achieved,
remains unknown at present. 

\begin{acknowledgments}
This work benefited from support through Project Nos. GR 967/7-1 and
GI 1405/1-1 of the Deutsche Forschungsgesellschaft (DFG). The authors
gratefully acknowledge the Gauss Centre for 
Supercomputing e.V. (www.gauss-centre.eu) for funding this project by
providing computing time through the John von Neumann Institute for
Computing (NIC) on the GCS Supercomputer 
JUWELS at J{\"u}lich Supercomputing Centre (JSC). J{\'a}n
{\v{S}}imkanin is grateful for support from the Alexander-von-Humboldt
Stiftung (CZE 1079936) and for funding from the Helmholtz Association
in frame of the AI project GEOMAGFOR (ZT-I-PF-5-200). 
\end{acknowledgments}

\bibliography{biblio}% Produces the bibliography via BibTeX.

\end{document}